\documentclass[10pt,letterpaper]{article}
\usepackage[top=0.85in,left=2.75in,footskip=0.75in]{geometry}

\usepackage{amsmath,amssymb}

\usepackage{changepage}

\usepackage{textcomp,marvosym}

\usepackage{cite}

\usepackage{nameref,hyperref}

\usepackage[table]{xcolor}

\usepackage{array}

\newcolumntype{+}{!{\vrule width 2pt}}

\newlength\savedwidth

\raggedright
\usepackage[aboveskip=1pt,labelfont=bf,labelsep=period,justification=raggedright,singlelinecheck=off]{caption}

\makeatletter
\renewcommand{\@biblabel}[1]{\quad#1.}
\makeatother

\usepackage{lastpage,fancyhdr,graphicx}
\usepackage{epstopdf}
\fancyheadoffset[L]{2.25in}
\fancyfootoffset[L]{2.25in}
\usepackage[capitalize]{cleveref}
\usepackage{xcolor}
\newcommand{\J}{\color{black} }

\begin{document}
\vspace*{0.2in}

\begin{flushleft}
{\Large
\textbf\newline{Trust based attachment} 
}
\newline
\\
Julian Kates-Harbeck\textsuperscript{1},
Martin A. Nowak\textsuperscript{2,3*},
\\
\bigskip

\textbf{1} Department of Physics,
\\
\textbf{2} Department of Mathematics,
\\
\textbf{3} Department of Organismic and Evolutionary Biology, Harvard University, Cambridge MA~02138,~USA
\\
\bigskip

* martin\_nowak@harvard.edu 

\end{flushleft}
\section*{Abstract}
In social systems subject to indirect reciprocity, a positive reputation is key for increasing one's likelihood of future positive interactions \cite{alexander1987biology,nowak1998evolution,wedekind2000cooperation,ohtsuki2004should,brandt2005indirect,nowak2005evolution,chalub2006evolution},\cite{nowak2006five,engelmann2009indirect,sigmund2010calculus,berger2011learning,radzvilavicius2018evolution,raub1990reputation}.
The flow of gossip can amplify the impact of a person's actions on their reputation depending on how widely it spreads across the social network, which leads to a percolation problem \cite{PhysRevE.76.036117}. To quantify this notion, we calculate the expected number of individuals, the ``audience'', who find out about a particular interaction. For a potential donor, a larger audience constitutes higher reputational stakes, and thus a higher incentive, to perform ``good'' actions in line with current social norms\cite{chalub2006evolution,santos2018social}. For a receiver, a larger audience therefore increases the trust that the partner will be cooperative. 
This idea can be used for an algorithm that generates social networks, which we call trust based attachment (TBA). TBA produces graphs that share crucial quantitative properties with real-world networks, such as high clustering, small-world behavior, and powerlaw degree distributions\cite{watts1998collective,barabasi1999emergence,easley2010networks,papadopoulos2012popularity,apicella2012social,pasquaretta2014social}. We also show that TBA can be approximated by simple friend-of-friend routines based on triadic closure, which are known to be highly effective at generating realistic social network structures \cite{kolda2014scalable,papadopoulos2012popularity,bhat2014emergence,lambiotte2016structural,leskovec2005graphs}. Therefore, our work provides a new justification for triadic closure in social contexts based on notions of trust, gossip, and social information spread. These factors are thus identified as potential significant influences on how humans form social ties.


{\J
\section*{Introduction}
Models of indirect reciprocity study the effect of reputational information on the actions of individuals in a group setting. They typically assume well mixed population structures \cite{alexander1987biology,chalub2006evolution,wedekind2000cooperation,sigmund2010calculus,engelmann2009indirect,santos2018social,nowak1998evolution,brandt2005indirect, nowak2005evolution,ohtsuki2004should,berger2011learning,radzvilavicius2018evolution}. However, because such information spreads via communication among individuals, it is affected by the structure of the social network underlying the group. A more sophisticated model of indirect reciprocity therefore might also take into account the effects of network structure \cite{mohtashemi2003evolution,wu2017influence}.  The spread of information on networks generally (including effects of social contagion), as well as its dependendence on global and local network features has been explored extensively \cite{bakshy2012role,montanari2010spread,centola2018behavior,wu2008community,keeling2011insights,borge2013cascading,watts2002simple}. However, the influence of network structural effects on reputational information spread specifically remains poorly studied.

While it is clear that network structure can influence reputation spread, the reverse is also true. The connections between individuals themselves can develop dynamically based on choices and incentives shaped by indirect reciprocity \cite{pujol2005can,vega2006building,traag2011indirect,corten2014computational,mohtashemi2003evolution}. In general, this leads to a complex interdependence between the spread of reputational information and network structure, potentially across multiple time scales, which poses a formidable modeling challenge. As a step towards jointly modeling reciprocity and network evolution, some papers have proposed models including information flow and link formation on networks in the context of repeated games \cite{pujol2005can,vega2006building,traag2011indirect,corten2014computational,mohtashemi2003evolution}. In these models, an individual may update links based on the past behavior of other agents in the repeated game, resulting in networks that co-evolve with individuals' strategies. While these papers do not study directly which network properties enhance or suppress the spread of gossip, their findings that cohesive, tightly clustered networks encourage cooperative behavior \cite{traag2011indirect,vega2006building} are in line with our results. These models do not consider strategic link formation or the effects of separate time scales between network evolution and individual behaviors.

Instead of attempting to capture the full interdependence of reputational and structural effects, we propose here an intermediary step where we first identify local structural features that relate strongly to reputation spread, and then study attachment strategies for individuals interested in building cooperative relationships in the presence of indirect reciprocity. Our objective is to provide a crisper mechanistic understanding than might be possible in a co-evolving model. We aim to illuminate static network features affecting reputation spread, as well as provide a possible game theoretical justification for some of the observed (growth) properties of real networks. The two key assumptions are (i) that reputation information can spread across the local network \cite{bakshy2012role,montanari2010spread,centola2018behavior,wu2008community,keeling2011insights,borge2013cascading,watts2002simple,PhysRevE.76.036117,johansson2017gossip} and (ii) that individuals seek to form attachments that are likely to be cooperative (either for forward looking strategic reasons or because this happens to be a winning social strategy in a repeated game setting) \cite{pujol2005can,vega2006building,traag2011indirect,corten2014computational}.  

In particular, we introduce a simple model of reputation spread that considers the expected number of individuals $n_{ij}$ learning about an actor $i$'s good deed to one of their neighbors $j$, as a function of the local network structure. The model has a single parameter $p$ which represents the strength or likelihood of communication and gossip spread between individuals. We take the quantity $n_{ij}$ to be a proxy of the strength of reputational incentives for that relationship. This motivates a strategy of attachment where individuals select for new links that increase incentives for mutually beneficial relationships. We call the resulting algorithm ``trust-based attachment'' (TBA). We show empirically and motivate mathematically that the simple and well-studied network formation strategy of triadic closure represents a realistic and effective heuristic for TBA in the limit of weak gossip (low $p$). Since triadic closure specifically generates network properties consistent with several features of real social networks (better even than the original TBA), we conclude that triadic closure may both be the real mechanism used in network growth, and also that it may have developed as a heuristic for a TBA-like attachment strategy. As such, we develop a link between the dynamics of indirect reciprocity and network growth. 

There is an extensive literature on models of network growth that can reproduce various statistical properties \cite{albert2002statistical,orsini2015quantifying} of real world networks. Small world behavior can arise with a small number of random long-range connections in otherwise densely locally connected networks \cite{watts1998collective}. Models based on randomly connected groups of communities of various sizes and densities \cite{kolda2014scalable} can reproduce both clustering and degree statistics of target networks. Scale free degree distributions can emerge from preferential attachment \cite{barabasi1999emergence}, and a trade-off between popularity and similarity can accurately model additional common network properties such as their hyperbolic geometry and high local clustering \cite{papadopoulos2012popularity}. 

While there is evidence for scale free degree distributions in real newtorks, as well as for preferential attachment-like affinities during network growth \cite{albert2002statistical}, it should be noted that not all networks with heavy tailed degree distributions follow true power-law behavior. Moreover, the mechanism underlying such properties can often provide more insight than their mere statistical observation \cite{clauset2009power,broido2019scale,stumpf2012critical}. More generally, while the above models have applicability and supporting evidence both in social and other (e.g. biological, technological, physical) contexts, the justification for their mechanisms in all those settings is not always clear. Several of these models also seemingly require global knowledge of the network (such as the degree of all nodes) for an agent to effectively implement their attachment mechanisms. One of our goals therefore is to provide mechanistic justification for a network generation mechanism that is easily implementable with local knowledge, and that can explain several of the social network features considered important in prior work, including short mean path length \cite{watts1998collective}, heavy-tailed degree distributions \cite{barabasi1999emergence}, and high local clustering \cite{papadopoulos2012popularity}. 

Triadic closure, i.e. forming a new link to a neighbor of a neighbor on the network, is an agent based, local mechanism common in realistic social settings \cite{bhat2014emergence,davidsen2002emergence,leskovec2005graphs}. It is known that ``trust'' in the resulting relationship and the associated ``social capital'' form a rationale for the appearance of triadic closure in social contexts (among opportunity and incentive for removing conflict) \cite{easley2010networks,raub2013rationality}, although the diffusion of information as a function of network structure has not been modeled explicitly in this context ($p$ is taken as $1$). Generative network models based on triadic closure with only few additional parameters can reproduce the characteristics of real social networks quantitatively \cite{bhat2014emergence,wu2015emergent,bhat2016densification,bianconi2014triadic,lambiotte2016structural,leskovec2005graphs,jackson2007meeting}. In this work, we do not aim to provide a more realistic algorithm for network growth, or one that more closely reproduces or fits real networks than existing work. Instead, we motivate TBA as a network growth mechanism based on indirect reciprocity, and then show that it is well approximated by triadic closure, which is easily implementable by agents with only local knowledge, and well studied in the literature as a realistic mechanism for network growth. Several properties of real social networks and their growth are not captured, including for example effects of homophily, node heterogeneity, or spatial constraints \cite{holzhauer2013considering,kim2017effect,avin2020mixed,murase2019structural}.

We provide further review of related work and the relationship to this paper in \nameref{S1_SI}. The rest of the main text is organized as follows: we first describe and study our proposed model of reputational information spread. Later, we use the results to motivate trust-based attachment. Finally, we study the networks generated by TBA and show that triadic closure forms a simple and practical heuristic for TBA.
}

\section*{Model of reputation spread}
\begin{figure*}[p]
\noindent\makebox[\linewidth]{
  \includegraphics[width=1.0\linewidth]{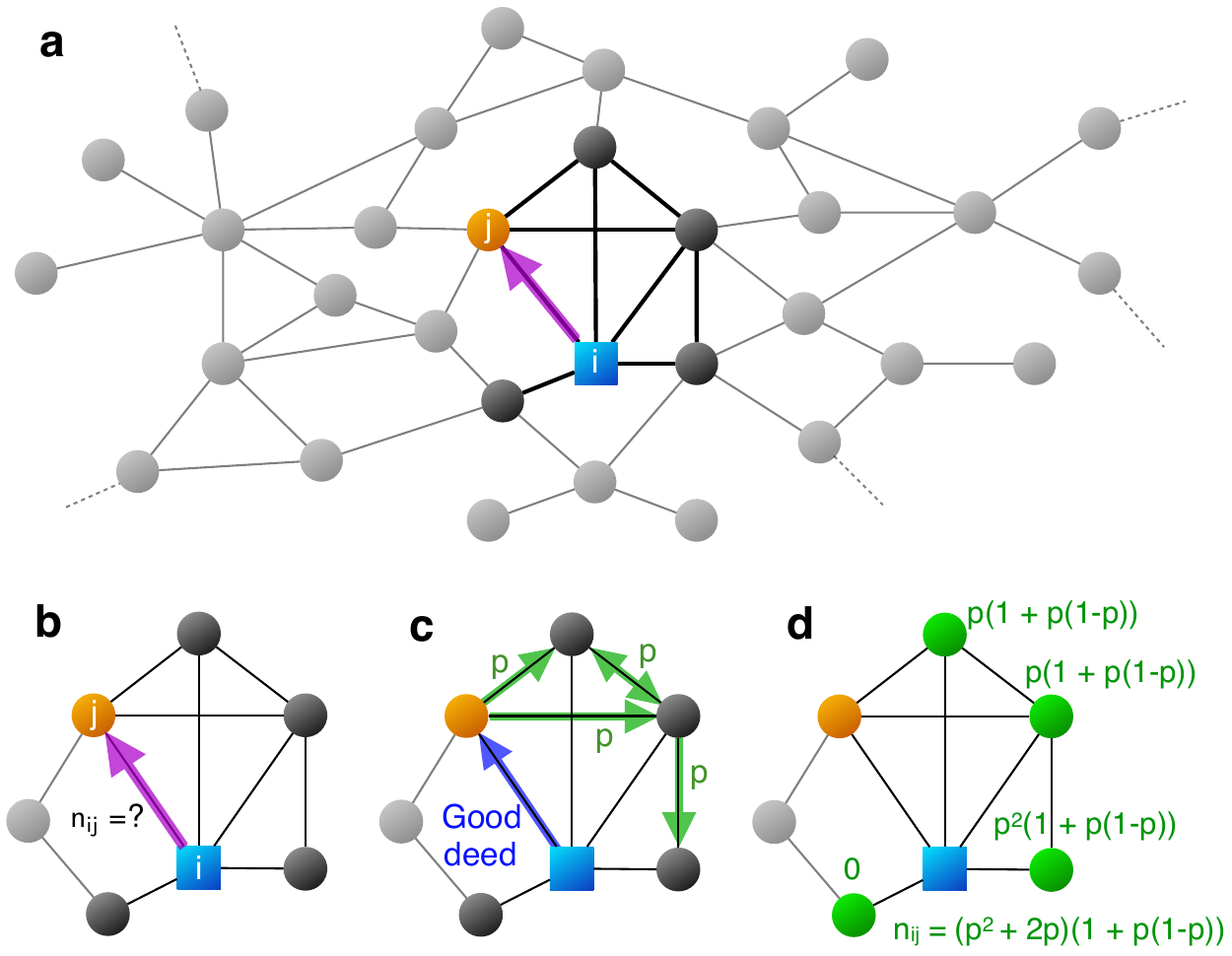}
}
\caption{
{\bf Calculating the size of the audience of a good deed.}
 {\bf a,} Individual $i$ (blue square) is connected to individual $j$ (orange circle) on a social network. {\bf b,} If $i$ performs a cooperative act toward $j$, who among the neighborhood of $i$ (dark nodes and edges) will find out about it? {\bf c,} We assume that gossip originates from the recipient, $j$, and flows with probability $p$ to other individuals as long as they know the donor, $i$.{\bf d,} Depending on the structure of the network, individuals have certain probabilities  to learn about the cooperative act (shown in green). Summing over these probabilities gives the expected number of individuals $n_{ij}$, which is the size of the audience for an action from $i$ to $j$.}

\label{fig:game_description}
\end{figure*}

Consider a population of individuals occupying the nodes of a graph. The edges determine mutual acquaintance, which implies possibilities of social interaction and communication \cite{sommerfeld2007gossip}. On the background of this social structure we consider a  game of gossip and social information spread (\cref{fig:game_description} ). An individual, the ``actor'', has the option to perform a cooperative act for one of her neighbors, the ``recipient''. Other individuals learn about the good deed in the following way: the gossip originates from the recipient {\J (we do not consider here the effect of the originator directly advertising their own action)}; each individual, who knows about the good deed, transmits the information to each of its neighbors with probability $p$. We assume that information can only flow between two individuals who both know the actor \cite{PhysRevE.76.036117}. {\J In \nameref{S1_SI} section ``Global spread'' we study the consequences of relaxing this constraint: as the probability of gossip spreading beyond neighbors of $i$ increases from $0$, global structural properties of the network become increasingly important.}

For a given graph, our model has a single parameter, $p$, the probability of gossip transfer over an edge. The basic calculation that we perform is the following: for a given donor $i$ and recipient $j$ on the graph, what is the expected number, $n_{ij}$, of third party individuals that learn about the cooperative act {\J with information originating from the recipient $j$}. The quantity $n_{ij}$ is the expected size of the {\J reputational} audience of $i$'s action toward $j$. The larger the value of $n_{ij}$ the more incentive  there is - based on considerations of reputation and indirect reciprocity \cite{alexander1987biology,nowak2005evolution} - for a cooperative act from $i$ to $j$, both due to the threat of  punishment for uncooperative behavior and the possibility of reward for cooperative behavior. In interpreting $n_{ij}$, we explicitly assume that there will be sufficient upcoming future interactions to make these considerations of the future relevant/dominant.

We have 
$$n_{ij}=\sum_{k \neq j} P_{jk} \eqno(1)$$
The index $k$ runs over all neighbors of the individual $i$, but omitting $j$. The quantity $P_{jk}$ denotes the probability that gossip \cite{PhysRevE.76.036117} originating from $j$ reaches $k$. $P_{jk}$ grows with the number of paths that can carry information, but falls off with their length. Note that  $P_{jk}$ is the percolation probability with parameter $p$ between node $j$ and $k$  on the sub-graph that is given by neighborhood of individual $i$. Calculating percolation on general graphs is a well-known, formidable problem, and  efficient algorithms exist only for special cases \cite{gilbert1959random,kirkpatrick1973percolation,PhysRevE.76.036117,PhysRevLett.85.5468}. However, in \nameref{S1_SI} section ``Random neighborhood approximation for gossip propagation'' we derive a fast and accurate approximation for calculating $n_{ij}$ as a function of the local degree, clustering, and embeddedness by leveraging the reliability theory of random graphs \cite{gilbert1959random}. We test the  applicability of this approximation by comparing it to exact calculations on small graphs and to extensive numerical simulations of both real-world and artificial networks (Extended Data Figures 1 and 2 in \nameref{S1_SI}).

\begin{figure*}[p]
\noindent\makebox[\linewidth]{
  \includegraphics[width=1.0\linewidth]{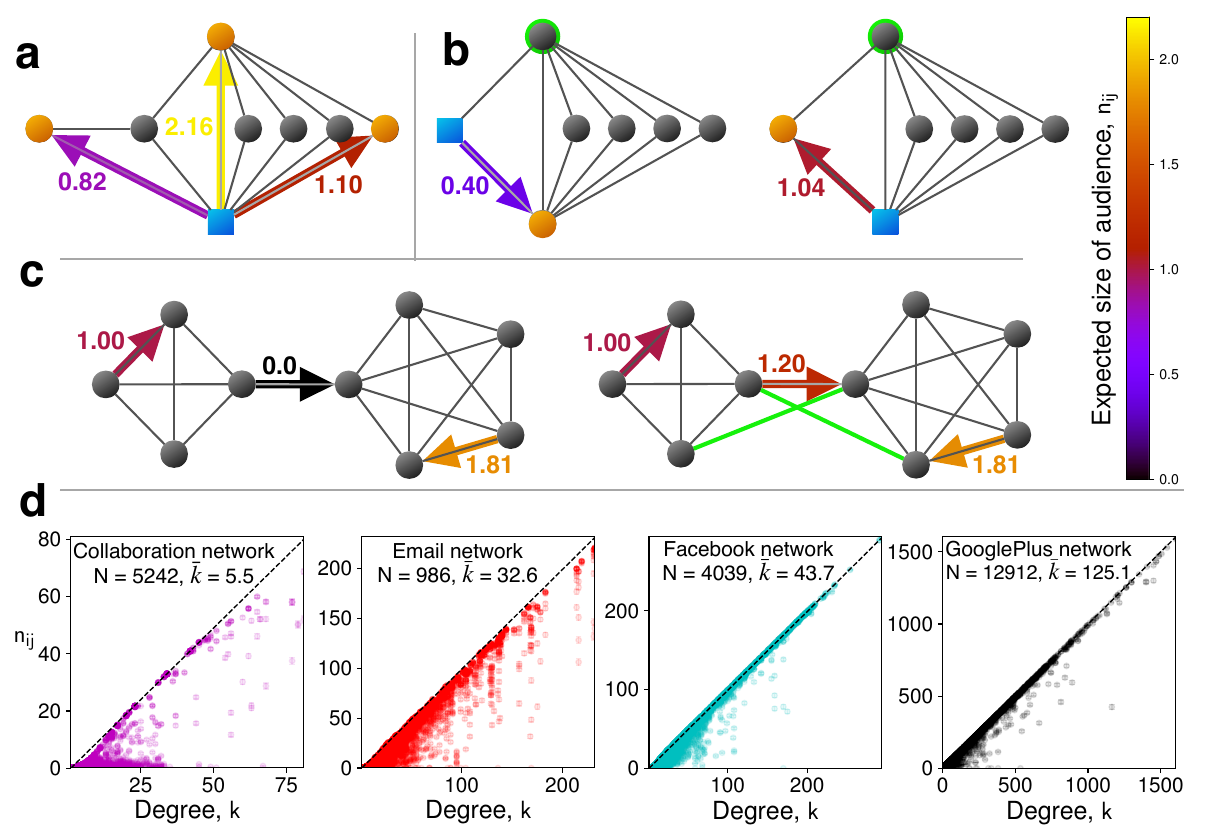}
}
\caption{{\bf Properties of $n_{ij}$ on local structures and real-world networks.} Values of $n_{ij}$ are shown as the colors of the respective arrows. {\bf a,} the value of $n_{ij}$ varies between recipients (orange circles) of the same actor (blue square). The coloring is as in \cref{fig:game_description}. If the recipient, $j$, is central to the neighborhood of the actor, $i$, and connected to several others, then $n_{ij}$ is large (yellow arrow). For a recipient, who is only peripherally connected, $n_{ij}$ is lower (red arrow). 
{\bf b,} {\J the value of $n_{ij}$ can differ from $n_{ji}$. In both cases, the overall networks are identical and the actor and recipient have a single mutual neighbor in common (highlighted in green). In the left case, the mutual neighbor is the actor's only additional neighbor, and $n_{ij}$ is low. In the right case, the actor has several other neighbors who can learn about the interaction, and $n_{ij}$ is higher.
{\bf c,} the value of $n_{ij}$ is large between nodes that are part of the same highly connected community (magenta and orange arrows). Two such communities may only be weakly connected to each other, and thus $n_{ij}$ is low for inter-community interactions. By adding some new links (thick green edges), higher incentives to cooperate (red arrow) can be established. Enhancing the interconnectivity between communities builds trust. Parameters: $p = 0.4$. For clarity we here omit coloring actors and recipients separately.
{\bf d,} Values for $n_{ij}$ are shown for $1000$ random edges on various real social networks \cite{snapnets,leskovec2007graph,leskovec2012learning} (see \nameref{S1_SI} for details).  Highly connected nodes have values of $n_{ij}$ near the maximum possible $k-1$ (see ``Size of the expected audience'' in \nameref{S1_SI}; black dashed diagonal). The total number of nodes $N$ and the mean degree $\bar{k}$ are given for each network.}
Parameters: $p = 0.2$.
}
\label{fig:P_neighborhood_examples}
\end{figure*}

In \cref{fig:P_neighborhood_examples}, we show various examples of network structures and make the following observations. (i) For a given actor, $n_{ij}$ can vary strongly between recipients. Links to individuals who are weakly connected with or even isolated from the rest of the actor's neighbors receive only small values of $n_{ij}$. By contrast, links to neighbors that are centrally located and highly connected to many other neighbors carry large values of $n_{ij}$. Such individuals can distribute information widely and thereby generate large effective audience sizes. (ii) The embeddedness \cite{watts1998collective,easley2010networks} of edge $ij$, which is the number of mutual neighbors between nodes $i$ and $j$, has has a strong predictive influence on the value of $n_{ij}$ (see Extended Data Figures 1 and 2 as well as ``Size of the expected audience'' in \nameref{S1_SI}). (iii) The value $n_{ij}$ can be very different from $n_{ji}$. Some links can have good incentives for cooperation in one direction but not in the other. (iv) The value $n_{ij}$ can be changed dramatically by modifying just a few key connections. (v) Overall, having actor and recipient be part of a large, densely interconnected community results in the largest values of $n_{ij}$. 

In \cref{fig:P_neighborhood_examples}, we also illustrate the behavior of $n_{ij}$ on four real-world social networks. These networks represent professional (``Collaboration'' and ``Email'') \cite{leskovec2007graph} and personal (``Facebook'' and ``Google Plus'') \cite{leskovec2012learning} relationships comprising thousands of individuals, each of which with tens to hundreds of links. We find that well connected nodes universally have high values of $n_{ij}$ towards their neighbors. On real social networks, gossip almost surely percolates throughout sufficiently large neighborhoods \cite{PhysRevE.76.036117}. 

For every directed edge on a social network, the parameter $n_{ij}$ is the expected size of the audience who find out about a cooperative act from $i$ to $j$. Thus, $n_{ij}$ determines the incentive  for individual $i$ to be cooperative toward individual $j$ based on reputational consequences. Conversely, $n_{ij}$ also determines the trust of $j$ that $i$ will cooperate \cite{christakis2010empirical}. While various notions of ``trust'' have been defined \cite{coleman1994foundations}, here we use the word in the particular sense that we expect that another player will be cooperative. 

In \nameref{S1_SI}, we also extend our model to the case of ``global cooperation'', where an individual has the option to perform a cooperative act such as a public service or a donation that is not directed towards any particular recipient. In that case, the act may be observed by any of the individuals' neighbors and gossip can then originate (see the  section ``Global cooperation'' and Extended Data Figures 3 to 5 in \nameref{S1_SI}). 

\section*{Trust-based attachment}
\begin{figure*}[p]
\noindent\makebox[\linewidth]{
  \includegraphics[width=1.0\linewidth]{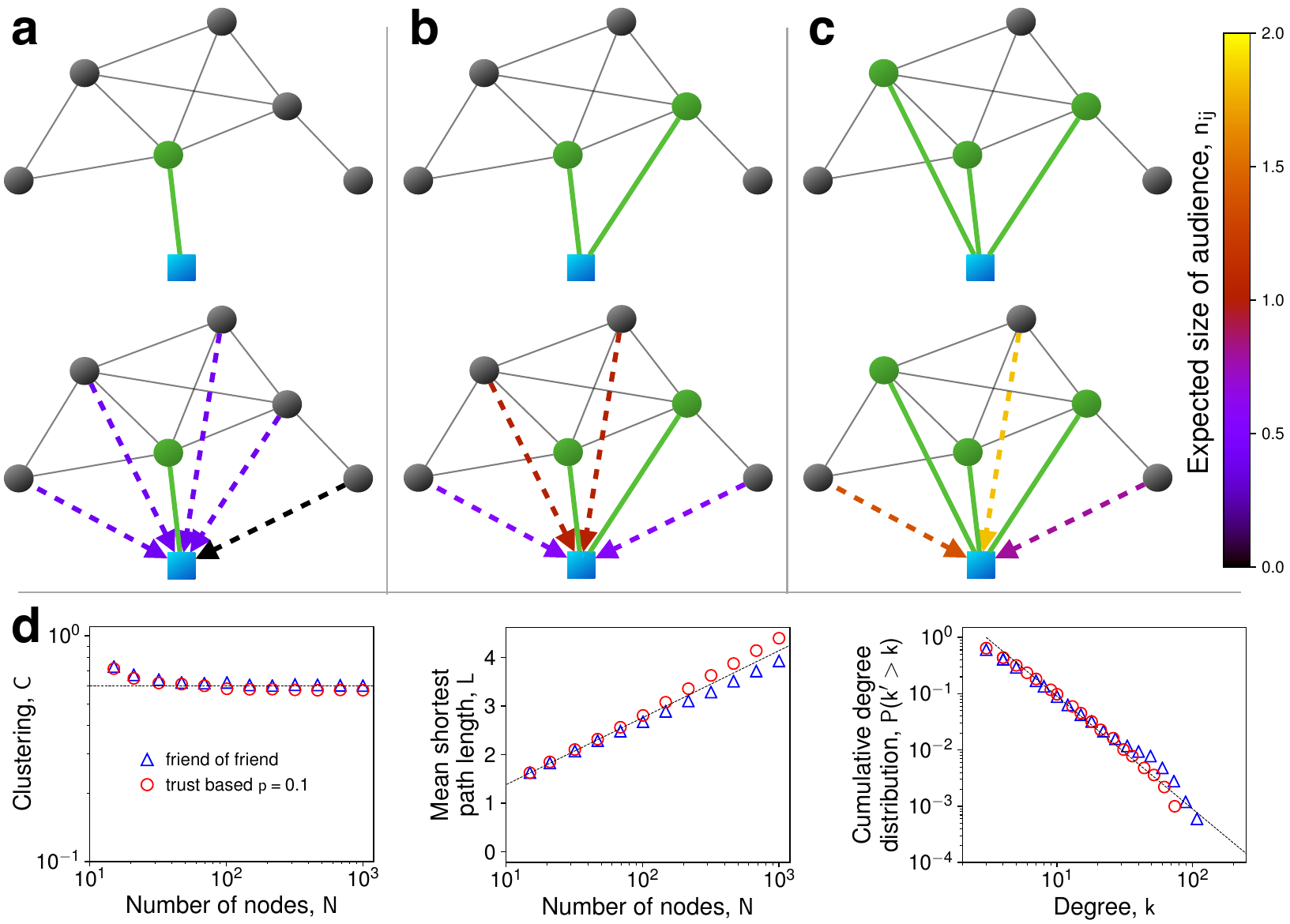}
}

\caption{
{\bf Trust based attachment}. The attachment process {\bf(a-c)} begins by a new individual (blue square) being introduced to a social group by a random individual (green) of that group. The new individual $j$ selects additional friends $k$ proportional to the trustworthiness of that link --- that is the value $n_{kj}$ --- if it were formed. The broken lines, in color, indicate the values of $n_{kj}$ on the edges that could be formed. Every time the new individual actually forms a link, the values of $n_{kj}$ for future potential friends change. The new individual in this case selects two additional friends for a total of three connections (green).
{\bf d,} comparing graphs generated by TBA (red circles) and by friend-of-friend attachment (blue triangles).  The TBA graphs display high and constant (as $N \to \infty$) clustering. The dashed line indicates the limiting value. The networks also show small world behavior, i.e. logarithmically growing mean shortest path length (the dashed line indicates $log N$ behavior), as well as a power law degree distribution (the dashed line has a slope of $-2$). Trust based attachment generates some of the same structural features as exhibited by real-world social networks, even in the simplified local form of friend-of-friend attachment. Error bars in both plots are significantly smaller than the size of the symbols. Parameters: $p = 0.1$, $k = 6$, $N \in [15,2000]$. For other values of $p$, see Extended Data Figure 6 in \nameref{S1_SI}.
}
\label{fig:trust_attachment_process}
\end{figure*}

We now use these insights to propose a mechanism for generating social networks  (\cref{fig:trust_attachment_process}). Consider a newcomer who is introduced to a social group by a random individual from that group. Thus, the newcomer's first link is formed to a random individual. Subsequently, the newcomer seeks to attach to individuals who have a high incentive to be cooperative toward her \cite{jackson2010social,easley2010networks,raub2013rationality}. This goal can be achieved in the following way. The newcomer, $j$, attaches to individual $i$ with a probability that is proportional to the quantity $n_{ij}$ on the potential link. The larger $n_{ij}$ the more incentive for $i$ to cooperate with $j$ and thus the more reason for $j$ to trust that $i$ is cooperative toward her. We assume here that the target individual for the new attachment $i$ only accepts newcomers with a given constant probability. As this does not influence the attachment statistics (but only the effective time scale of the process), we omit this influence in the following discussion.

In order to generate a network of $N$ total individuals with average degree $k$, we begin with a fully connected graph of $k+1$ individuals. Then we add individuals one at a time until there are $N$ individuals in total. Each new individual forms $k/2$ connections as described above. We propose to call this algorithm for generating social networks ``trust based attachment" (TBA). In this algorithm, trust forms the glue of social networks.

{\J Naturally, TBA forms a simplified model that does not capture all aspects underlying the formation of real-life social networks. Therefore, we do not expect this model to reproduce all features of real social networks. For example, it does not reproduce the distinct community structures commonly found in real social networks, possibly due to a lack of modeling homophily, spatial constraints, or heterogeneity in the properties of individuals \cite{holzhauer2013considering,kim2017effect,avin2020mixed,murase2019structural}.  It is also not an attempt to quantitatively ``fit'' the properties of real social networks. Instead, we aim to show that this simple mechanism generates several important properties of real social networks, thus providing a possible mechanistic justification for their emergence. In particular,} TBA generates networks that have several desirable, realistic features (\cref{fig:trust_attachment_process} and Extended Data Figure 6 in \nameref{S1_SI}), which include include high clustering, power law degree distributions, and small world behavior. 

\begin{figure*}[p]
\centering
\noindent\makebox[\linewidth]{
  \includegraphics[width=0.7\linewidth]{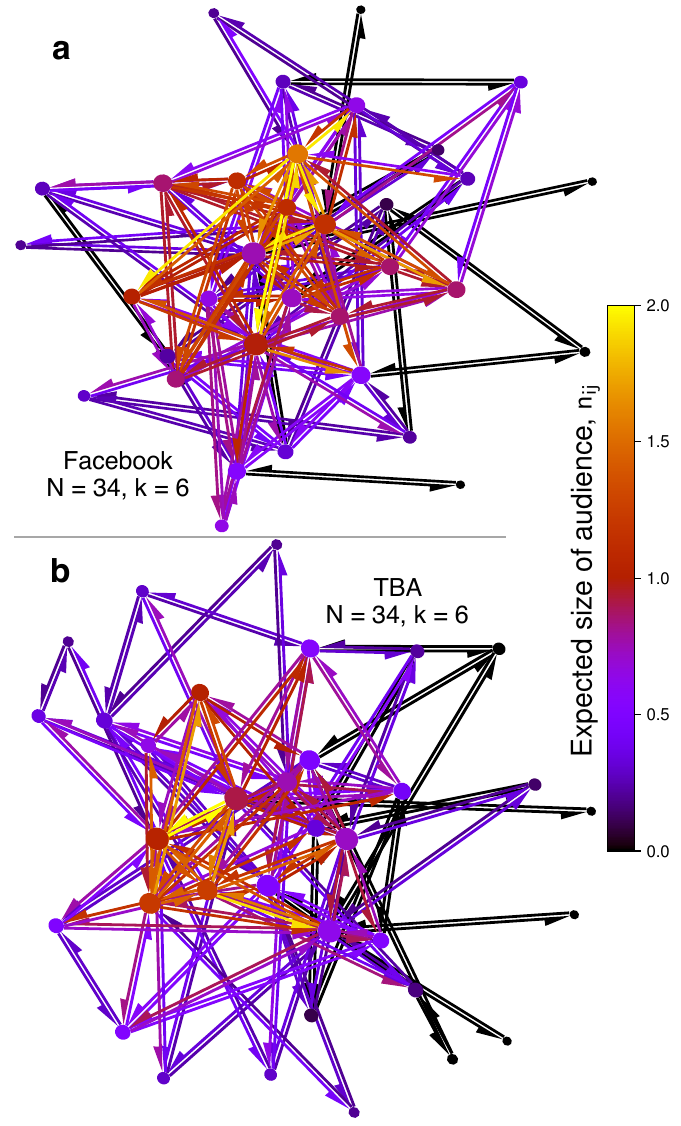}
}
\caption{{\bf Illustrating $n_{ij}$ on a Facebook and a TBA network.} A subset of a Facebook social network \cite{snapnets,leskovec2012learning} is shown at the top, and a network generated by TBA with the same number of nodes ($N$) and average degree ($k$) is shown below. Edges are colored by $n_{ij}$ and nodes are colored by the value of $n_{ij}$ averaged over all neighbors $j$, which we denote as $n_{i*}$ . While $n_{ij}$ is specific to the relationship from $i$ to $j$, $n_{i*}$ can be seen as a measure of ``average trustworthiness'' of the individual $i$. {\J While these networks are quite small, they allow a clearer view into the local structure. The networks have strongly connected central hubs and high clustering. The highest values of $n_{ij}$ appear with strongly embedded nodes in the center, while the lowest values appear with the isolated peripheral nodes.} Parameters: $p = 0.2$. See Extended Data Figures 7 and 8 in \nameref{S1_SI} for more details.}
\label{fig:P_real_examples}
\end{figure*}

{\J To build some further intuition for the key quantities and network models introduced in this paper, we show in \cref{fig:P_real_examples} a component of a real social network as well as an equivalent network of the same size and average degree generated by TBA. The purpose of this illustration is not to show quantitative agreement. Rather, it is to illustrate the qualitative structure of these networks, and show the dependence of $n_{ij}$ on local structural features, embedded in a more realistic context than in previous figures.} Edges are colored by $n_{ij}$ and nodes are colored by the value of $n_{ij}$ averaged over all neighbors $j$, which we denote as $n_{i*}$ . While $n_{ij}$ is specific to the relationship from $i$ to $j$, $n_{i*}$ can be seen as a measure of ``average trustworthiness'' of the individual $i$, and its statistics have been studied extensively on real-world as well as small world and preferential attachment networks \cite{PhysRevE.76.036117}. Even in the small networks in this figure one can see the varied degree distribution and the densely connected and highly clustered ``core''. For a more quantitative assessment of networks generated by TBA we refer to \cref{fig:trust_attachment_process}.

TBA captures {\J a subset of the qualitative features} of real social networks well, including high clustering, small world behavior, hubs, varied degree distributions and an overall high prevalence of trust\cite{watts1998collective,barabasi1999emergence,easley2010networks,papadopoulos2012popularity,apicella2012social,pasquaretta2014social,jackson2007meeting}. Nodes on the periphery of the social networks with only few attachments to distinct parts of the network have low values of $n_{i*}$. By contrast, highly connected nodes that are part of densely clustered communities display large trustworthiness.

\section*{Triadic closure as a plausible heuristic for TBA}
Implementing TBA in practice would require individuals to have extensive knowledge of the local network structure and to perform complicated calculations. However, we find (see \cref{fig:trust_attachment_process}, as well as Extended Data Figure 6 and ``Network generation'' in \nameref{S1_SI}) that the simple heuristic of attaching to a random neighbor of a random neighbor (``triadic closure'') can accurately approximate TBA for values of $p \lesssim 0.2$. This ``friend-of-friend'' attachment is simple and intuitive, and can be implemented by individual agents without global knowledge of the network structure \cite{pujol2005can}. Both methods generate networks with very similar connectivity statistics, including high clustering, short mean path length, and power law degree distributions (see \cref{fig:trust_attachment_process}). For triadic closure, the high clustering property emerges since it forms edges that are biased to be highly embedded, which means forming many triangles at once. The random shortening of paths with every additional edge creates the small world behavior. Attaching to a random neighbor of a node already involved in the network introduces a bias towards high degree nodes \cite{easley2010networks}, which reproduces the preferential attachment bias and thus power law degree distributions. For large values of $p > 0.2$ , we find that TBA still generates networks with nearly the same properties as friend-of-friend attachment, except that the degree distribution for TBA departs from a power law and includes even more high-degree ``hubs'' (see ``Network generation'' and Extended Data Figure 6 in \nameref{S1_SI}).  

{\J In summary, the incentives created by indirect reciprocity suggest triadic closure as a simple and practical heuristic for creating links with high trust. Triadic closure in turn can explain multiple characterstics of real social networks \cite{bhat2014emergence,davidsen2002emergence,leskovec2005graphs,bhat2014emergence,wu2015emergent,bhat2016densification,bianconi2014triadic,lambiotte2016structural,leskovec2005graphs,jackson2007meeting}. The key structural features connecting the spread of indirect reciprocity to triadic closure are related to the quantity $n_{ij}$, which in turn can be well approximated with knowledge of the local degree, clustering, and embeddedness for a given node. 

In principle, there are many ways to build social ties:  individuals could strive to form long range, random connections to increase diversity; seek out individuals of a certain type, status or connectivity; or look for meeting venues that bring together people with specific interests or motivations. While all these effects may play a role, our work helps explain why out of all the ways that humans could create new social connections, forming relationships with friends of friends is a natural, intuitive, and trust-inducing process, and one that is likely to generate trustworthy and cooperative connections.}

\section*{Figures}

\clearpage
\section*{Supporting Information}

\paragraph*{S1 SI.}
\label{S1_SI}
{\bf Supplemental Information.} This file contains supplemental discussions including sections with additional information on (i) the relationship with past work; (ii) details on methods and algorithms used; (ii) the calculation of the size of the expected audience; (iii) model assumptions and extensions; (iv) quantitative derivations around network generation; as well as (v) extended data figures.


\begin{thebibliography}{10}

\bibitem{alexander1987biology}
Alexander R.
\newblock The biology of moral systems (foundations of human behavior). 1987;.

\bibitem{nowak1998evolution}
Nowak MA, Sigmund K.
\newblock Evolution of indirect reciprocity by image scoring.
\newblock Nature. 1998;393(6685):573--577.

\bibitem{wedekind2000cooperation}
Wedekind C, Milinski M.
\newblock Cooperation through image scoring in humans.
\newblock Science. 2000;288(5467):850--852.

\bibitem{ohtsuki2004should}
Ohtsuki H, Iwasa Y.
\newblock How should we define goodness?—reputation dynamics in indirect
  reciprocity.
\newblock Journal of Theoretical Biology. 2004;231(1):107--120.

\bibitem{brandt2005indirect}
Brandt H, Sigmund K.
\newblock Indirect reciprocity, image scoring, and moral hazard.
\newblock Proceedings of the National Academy of Sciences.
  2005;102(7):2666--2670.

\bibitem{nowak2005evolution}
Nowak MA, Sigmund K.
\newblock Evolution of indirect reciprocity.
\newblock Nature. 2005;437(7063):1291--1298.

\bibitem{chalub2006evolution}
Chalub FA, Santos FC, Pacheco JM.
\newblock The evolution of norms.
\newblock Journal of theoretical biology. 2006;241(2):233--240.

\bibitem{nowak2006five}
Nowak MA.
\newblock Five rules for the evolution of cooperation.
\newblock Science. 2006;314(5805):1560--1563.

\bibitem{engelmann2009indirect}
Engelmann D, Fischbacher U.
\newblock Indirect reciprocity and strategic reputation building in an
  experimental helping game.
\newblock Games and Economic Behavior. 2009;67(2):399--407.

\bibitem{sigmund2010calculus}
Sigmund K.
\newblock The calculus of selfishness.
\newblock Princeton University Press; 2010.

\bibitem{berger2011learning}
Berger U.
\newblock Learning to cooperate via indirect reciprocity.
\newblock Games and Economic Behavior. 2011;72(1):30--37.

\bibitem{radzvilavicius2018evolution}
Radzvilavicius A, Stewart A, Plotkin JB.
\newblock Evolution of empathetic moral evaluation.
\newblock bioRxiv. 2018; p. 447151.

\bibitem{raub1990reputation}
Raub W, Weesie J.
\newblock Reputation and efficiency in social interactions: An example of
  network effects.
\newblock American Journal of Sociology. 1990; p. 626--654.

\bibitem{PhysRevE.76.036117}
Lind PG, da~Silva LR, Andrade JS, Herrmann HJ.
\newblock Spreading gossip in social networks.
\newblock Phys Rev E. 2007;76:036117.
\newblock doi:{10.1103/PhysRevE.76.036117}.

\bibitem{santos2018social}
Santos FP, Santos FC, Pacheco JM.
\newblock Social norm complexity and past reputations in the evolution of
  cooperation.
\newblock Nature. 2018;555(7695):242.

\bibitem{watts1998collective}
Watts DJ, Strogatz SH.
\newblock Collective dynamics of ‘small-world’ networks.
\newblock Nature. 1998;393(6684):440--442.

\bibitem{barabasi1999emergence}
Barab{\'a}si AL, Albert R.
\newblock Emergence of scaling in random networks.
\newblock Science. 1999;286(5439):509--512.

\bibitem{easley2010networks}
Easley D, Kleinberg J.
\newblock Networks, crowds, and markets.
\newblock Cambridge Univ Press. 2010;6(1):1--6.

\bibitem{papadopoulos2012popularity}
Papadopoulos F, Kitsak M, Serrano M{\'A}, Bogun{\'a} M, Krioukov D.
\newblock Popularity versus similarity in growing networks.
\newblock Nature. 2012;489(7417):537--540.

\bibitem{apicella2012social}
Apicella CL, Marlowe FW, Fowler JH, Christakis NA.
\newblock Social networks and cooperation in hunter-gatherers.
\newblock Nature. 2012;481(7382):497.

\bibitem{pasquaretta2014social}
Pasquaretta C, Lev{\'e} M, Claidiere N, Van De~Waal E, Whiten A, MacIntosh AJ,
  et~al.
\newblock Social networks in primates: smart and tolerant species have more
  efficient networks.
\newblock Scientific reports. 2014;4:7600.

\bibitem{kolda2014scalable}
Kolda TG, Pinar A, Plantenga T, Seshadhri C.
\newblock A scalable generative graph model with community structure.
\newblock SIAM Journal on Scientific Computing. 2014;36(5):C424--C452.

\bibitem{bhat2014emergence}
Bhat U, Krapivsky PL, Redner S.
\newblock Emergence of clustering in an acquaintance model without homophily.
\newblock Journal of Statistical Mechanics: Theory and Experiment.
  2014;2014(11):P11035.

\bibitem{lambiotte2016structural}
Lambiotte R, Krapivsky P, Bhat U, Redner S.
\newblock Structural Transitions in Densifying Networks.
\newblock Physical review letters. 2016;117(21):218301.

\bibitem{leskovec2005graphs}
Leskovec J, Kleinberg J, Faloutsos C.
\newblock Graphs over time: densification laws, shrinking diameters and
  possible explanations.
\newblock In: Proceedings of the eleventh ACM SIGKDD international conference
  on Knowledge discovery in data mining. ACM; 2005. p. 177--187.

\bibitem{mohtashemi2003evolution}
Mohtashemi M, Mui L.
\newblock Evolution of indirect reciprocity by social information: the role of
  trust and reputation in evolution of altruism.
\newblock Journal of theoretical biology. 2003;223(4):523--531.

\bibitem{wu2017influence}
Wu H, Arenas A, G{\'o}mez S.
\newblock Influence of trust in the spreading of information.
\newblock Physical Review E. 2017;95(1):012301.

\bibitem{bakshy2012role}
Bakshy E, Rosenn I, Marlow C, Adamic L.
\newblock The role of social networks in information diffusion.
\newblock In: Proceedings of the 21st international conference on World Wide
  Web. ACM; 2012. p. 519--528.

\bibitem{montanari2010spread}
Montanari A, Saberi A.
\newblock The spread of innovations in social networks.
\newblock Proceedings of the National Academy of Sciences.
  2010;107(47):20196--20201.

\bibitem{centola2018behavior}
Centola D.
\newblock How behavior spreads: The science of complex contagions. vol.~3.
\newblock Princeton University Press; 2018.

\bibitem{wu2008community}
Wu X, Liu Z.
\newblock How community structure influences epidemic spread in social
  networks.
\newblock Physica A: Statistical Mechanics and its Applications.
  2008;387(2):623--630.

\bibitem{keeling2011insights}
House T, Keeling MJ.
\newblock Insights from unifying modern approximations to infections on
  networks.
\newblock Journal of The Royal Society Interface. 2011;8(54):67--73.

\bibitem{borge2013cascading}
Borge-Holthoefer J, Ba{\~n}os RA, Gonz{\'a}lez-Bail{\'o}n S, Moreno Y.
\newblock Cascading behaviour in complex socio-technical networks.
\newblock Journal of Complex Networks. 2013;1(1):3--24.

\bibitem{watts2002simple}
Watts DJ.
\newblock A simple model of global cascades on random networks.
\newblock Proceedings of the National Academy of Sciences.
  2002;99(9):5766--5771.

\bibitem{pujol2005can}
Pujol JM, Flache A, Delgado J, Sang{\"u}esa R.
\newblock How can social networks ever become complex? Modelling the emergence
  of complex networks from local social exchanges.
\newblock Journal of Artificial Societies and Social Simulation. 2005;8(4).

\bibitem{vega2006building}
Vega-Redondo F.
\newblock Building up social capital in a changing world.
\newblock Journal of Economic Dynamics and Control. 2006;30(11):2305--2338.

\bibitem{traag2011indirect}
Traag VA, Van~Dooren P, Nesterov Y.
\newblock Indirect reciprocity through gossiping can lead to cooperative
  clusters.
\newblock In: Artificial Life (ALIFE), 2011 IEEE Symposium on. IEEE; 2011. p.
  154--161.

\bibitem{corten2014computational}
Corten R.
\newblock Computational Approaches to Studying the Co-evolution of Networks and
  Behavior in Social Dilemmas.
\newblock John Wiley \& Sons; 2014.

\bibitem{johansson2017gossip}
Johansson T.
\newblock Gossip spread in social network Models.
\newblock Physica A: Statistical Mechanics and its Applications.
  2017;471:126--134.

\bibitem{albert2002statistical}
Albert R, Barab{\'a}si AL.
\newblock Statistical mechanics of complex networks.
\newblock Reviews of modern physics. 2002;74(1):47.

\bibitem{orsini2015quantifying}
Orsini C, Dankulov MM, Colomer-de Sim{\'o}n P, Jamakovic A, Mahadevan P, Vahdat
  A, et~al.
\newblock Quantifying randomness in real networks.
\newblock Nature communications. 2015;6:8627.

\bibitem{clauset2009power}
Clauset A, Shalizi CR, Newman ME.
\newblock Power-law distributions in empirical data.
\newblock SIAM review. 2009;51(4):661--703.

\bibitem{broido2019scale}
Broido AD, Clauset A.
\newblock Scale-free networks are rare.
\newblock Nature communications. 2019;10(1):1017.

\bibitem{stumpf2012critical}
Stumpf MP, Porter MA.
\newblock Critical truths about power laws.
\newblock Science. 2012;335(6069):665--666.

\bibitem{davidsen2002emergence}
Davidsen J, Ebel H, Bornholdt S.
\newblock Emergence of a small world from local interactions: Modeling
  acquaintance networks.
\newblock Physical Review Letters. 2002;88(12):128701.

\bibitem{raub2013rationality}
Raub W, Buskens V, Frey V.
\newblock The rationality of social structure: Cooperation in social dilemmas
  through investments in and returns on social capital.
\newblock Social Networks. 2013;35(4):720--732.

\bibitem{wu2015emergent}
Wu Z, Menichetti G, Rahmede C, Bianconi G.
\newblock Emergent complex network geometry.
\newblock Scientific reports. 2015;5.

\bibitem{bhat2016densification}
Bhat U, Krapivsky P, Lambiotte R, Redner S.
\newblock Densification and structural transitions in networks that grow by
  node copying.
\newblock Physical Review E. 2016;94(6):062302.

\bibitem{bianconi2014triadic}
Bianconi G, Darst RK, Iacovacci J, Fortunato S.
\newblock Triadic closure as a basic generating mechanism of communities in
  complex networks.
\newblock Physical Review E. 2014;90(4):042806.

\bibitem{jackson2007meeting}
Jackson MO, Rogers BW.
\newblock Meeting strangers and friends of friends: How random are social
  networks?
\newblock American Economic Review. 2007;97(3):890--915.

\bibitem{holzhauer2013considering}
Holzhauer S, Krebs F, Ernst A.
\newblock Considering baseline homophily when generating spatial social
  networks for agent-based modelling.
\newblock Computational and Mathematical Organization Theory. 2013;19:128--150.

\bibitem{kim2017effect}
Kim K, Altmann J.
\newblock Effect of homophily on network formation.
\newblock Communications in Nonlinear Science and Numerical Simulation.
  2017;44:482--494.

\bibitem{avin2020mixed}
Avin C, Daltrophe H, Keller B, Lotker Z, Mathieu C, Peleg D, et~al.
\newblock Mixed preferential attachment model: Homophily and minorities in
  social networks.
\newblock Physica A: Statistical Mechanics and its Applications.
  2020;555:124723.

\bibitem{murase2019structural}
Murase Y, Jo HH, T{\"o}r{\"o}k J, Kert{\'e}sz J, Kaski K.
\newblock Structural transition in social networks: The role of homophily.
\newblock Scientific reports. 2019;9(1):1--8.

\bibitem{sommerfeld2007gossip}
Sommerfeld RD, Krambeck HJ, Semmann D, Milinski M.
\newblock Gossip as an alternative for direct observation in games of indirect
  reciprocity.
\newblock Proceedings of the national academy of sciences.
  2007;104(44):17435--17440.

\bibitem{gilbert1959random}
Gilbert EN.
\newblock Random graphs.
\newblock The Annals of Mathematical Statistics. 1959;30(4):1141--1144.

\bibitem{kirkpatrick1973percolation}
Kirkpatrick S.
\newblock Percolation and conduction.
\newblock Reviews of modern physics. 1973;45(4):574.

\bibitem{PhysRevLett.85.5468}
Callaway DS, Newman MEJ, Strogatz SH, Watts DJ.
\newblock Network Robustness and Fragility: Percolation on Random Graphs.
\newblock Phys Rev Lett. 2000;85:5468--5471.
\newblock doi:{10.1103/PhysRevLett.85.5468}.

\bibitem{leskovec2007graph}
Leskovec J, Kleinberg J, Faloutsos C.
\newblock Graph evolution: Densification and shrinking diameters.
\newblock ACM transactions on Knowledge Discovery from Data (TKDD).
  2007;1(1):2--es.

\bibitem{leskovec2012learning}
Leskovec J, Mcauley J.
\newblock Learning to discover social circles in ego networks.
\newblock Advances in neural information processing systems. 2012;25.

\bibitem{christakis2010empirical}
Christakis NA, Fowler JH, Imbens GW, Kalyanaraman K.
\newblock An empirical model for strategic network formation.
\newblock National Bureau of Economic Research; 2010.

\bibitem{coleman1994foundations}
Coleman JS.
\newblock Foundations of social theory.
\newblock Harvard university press; 1994.

\bibitem{jackson2010social}
Jackson MO.
\newblock Social and economic networks.
\newblock Princeton university press; 2010.

\bibitem{snapnets}
Leskovec J, Krevl A. {SNAP Datasets}: {Stanford} Large Network Dataset
  Collection; 2014.
\newblock \url{http://snap.stanford.edu/data}.

\end{thebibliography}


\begin{thebibliography}{10}

\bibitem{alexander1987biology}
Alexander R.
\newblock The biology of moral systems (foundations of human behavior). 1987;.

\bibitem{chalub2006evolution}
Chalub FA, Santos FC, Pacheco JM.
\newblock The evolution of norms.
\newblock Journal of theoretical biology. 2006;241(2):233--240.

\bibitem{wedekind2000cooperation}
Wedekind C, Milinski M.
\newblock Cooperation through image scoring in humans.
\newblock Science. 2000;288(5467):850--852.

\bibitem{sigmund2010calculus}
Sigmund K.
\newblock The calculus of selfishness.
\newblock Princeton University Press; 2010.

\bibitem{engelmann2009indirect}
Engelmann D, Fischbacher U.
\newblock Indirect reciprocity and strategic reputation building in an
  experimental helping game.
\newblock Games and Economic Behavior. 2009;67(2):399--407.

\bibitem{santos2018social}
Santos FP, Santos FC, Pacheco JM.
\newblock Social norm complexity and past reputations in the evolution of
  cooperation.
\newblock Nature. 2018;555(7695):242.

\bibitem{nowak1998evolution}
Nowak MA, Sigmund K.
\newblock Evolution of indirect reciprocity by image scoring.
\newblock Nature. 1998;393(6685):573--577.

\bibitem{brandt2005indirect}
Brandt H, Sigmund K.
\newblock Indirect reciprocity, image scoring, and moral hazard.
\newblock Proceedings of the National Academy of Sciences.
  2005;102(7):2666--2670.

\bibitem{nowak2005evolution}
Nowak MA, Sigmund K.
\newblock Evolution of indirect reciprocity.
\newblock Nature. 2005;437(7063):1291--1298.

\bibitem{ohtsuki2004should}
Ohtsuki H, Iwasa Y.
\newblock How should we define goodness?—reputation dynamics in indirect
  reciprocity.
\newblock Journal of Theoretical Biology. 2004;231(1):107--120.

\bibitem{berger2011learning}
Berger U.
\newblock Learning to cooperate via indirect reciprocity.
\newblock Games and Economic Behavior. 2011;72(1):30--37.

\bibitem{radzvilavicius2018evolution}
Radzvilavicius A, Stewart A, Plotkin JB.
\newblock Evolution of empathetic moral evaluation.
\newblock bioRxiv. 2018; p. 447151.

\bibitem{bakshy2012role}
Bakshy E, Rosenn I, Marlow C, Adamic L.
\newblock The role of social networks in information diffusion.
\newblock In: Proceedings of the 21st international conference on World Wide
  Web. ACM; 2012. p. 519--528.

\bibitem{montanari2010spread}
Montanari A, Saberi A.
\newblock The spread of innovations in social networks.
\newblock Proceedings of the National Academy of Sciences.
  2010;107(47):20196--20201.

\bibitem{centola2018behavior}
Centola D.
\newblock How behavior spreads: The science of complex contagions. vol.~3.
\newblock Princeton University Press; 2018.

\bibitem{wu2008community}
Wu X, Liu Z.
\newblock How community structure influences epidemic spread in social
  networks.
\newblock Physica A: Statistical Mechanics and its Applications.
  2008;387(2):623--630.

\bibitem{keeling2011insights}
House T, Keeling MJ.
\newblock Insights from unifying modern approximations to infections on
  networks.
\newblock Journal of The Royal Society Interface. 2011;8(54):67--73.

\bibitem{borge2013cascading}
Borge-Holthoefer J, Ba{\~n}os RA, Gonz{\'a}lez-Bail{\'o}n S, Moreno Y.
\newblock Cascading behaviour in complex socio-technical networks.
\newblock Journal of Complex Networks. 2013;1(1):3--24.

\bibitem{watts2002simple}
Watts DJ.
\newblock A simple model of global cascades on random networks.
\newblock Proceedings of the National Academy of Sciences.
  2002;99(9):5766--5771.

\bibitem{PhysRevE.76.036117}
Lind PG, da~Silva LR, Andrade JS, Herrmann HJ.
\newblock Spreading gossip in social networks.
\newblock Phys Rev E. 2007;76:036117.
\newblock doi:{10.1103/PhysRevE.76.036117}.

\bibitem{johansson2017gossip}
Johansson T.
\newblock Gossip spread in social network Models.
\newblock Physica A: Statistical Mechanics and its Applications.
  2017;471:126--134.

\bibitem{shaw2011effect}
Shaw AK, Tsvetkova M, Daneshvar R.
\newblock The effect of gossip on social networks.
\newblock Complexity. 2011;16(4):39--47.

\bibitem{pujol2005can}
Pujol JM, Flache A, Delgado J, Sang{\"u}esa R.
\newblock How can social networks ever become complex? Modelling the emergence
  of complex networks from local social exchanges.
\newblock Journal of Artificial Societies and Social Simulation. 2005;8(4).

\bibitem{vega2006building}
Vega-Redondo F.
\newblock Building up social capital in a changing world.
\newblock Journal of Economic Dynamics and Control. 2006;30(11):2305--2338.

\bibitem{traag2011indirect}
Traag VA, Van~Dooren P, Nesterov Y.
\newblock Indirect reciprocity through gossiping can lead to cooperative
  clusters.
\newblock In: Artificial Life (ALIFE), 2011 IEEE Symposium on. IEEE; 2011. p.
  154--161.

\bibitem{corten2014computational}
Corten R.
\newblock Computational Approaches to Studying the Co-evolution of Networks and
  Behavior in Social Dilemmas.
\newblock John Wiley \& Sons; 2014.

\bibitem{orsini2015quantifying}
Orsini C, Dankulov MM, Colomer-de Sim{\'o}n P, Jamakovic A, Mahadevan P, Vahdat
  A, et~al.
\newblock Quantifying randomness in real networks.
\newblock Nature communications. 2015;6:8627.

\bibitem{watts1998collective}
Watts DJ, Strogatz SH.
\newblock Collective dynamics of ‘small-world’ networks.
\newblock Nature. 1998;393(6684):440--442.

\bibitem{kolda2014scalable}
Kolda TG, Pinar A, Plantenga T, Seshadhri C.
\newblock A scalable generative graph model with community structure.
\newblock SIAM Journal on Scientific Computing. 2014;36(5):C424--C452.

\bibitem{barabasi1999emergence}
Barab{\'a}si AL, Albert R.
\newblock Emergence of scaling in random networks.
\newblock Science. 1999;286(5439):509--512.

\bibitem{papadopoulos2012popularity}
Papadopoulos F, Kitsak M, Serrano M{\'A}, Bogun{\'a} M, Krioukov D.
\newblock Popularity versus similarity in growing networks.
\newblock Nature. 2012;489(7417):537--540.

\bibitem{bhat2014emergence}
Bhat U, Krapivsky PL, Redner S.
\newblock Emergence of clustering in an acquaintance model without homophily.
\newblock Journal of Statistical Mechanics: Theory and Experiment.
  2014;2014(11):P11035.

\bibitem{davidsen2002emergence}
Davidsen J, Ebel H, Bornholdt S.
\newblock Emergence of a small world from local interactions: Modeling
  acquaintance networks.
\newblock Physical Review Letters. 2002;88(12):128701.

\bibitem{leskovec2005graphs}
Leskovec J, Kleinberg J, Faloutsos C.
\newblock Graphs over time: densification laws, shrinking diameters and
  possible explanations.
\newblock In: Proceedings of the eleventh ACM SIGKDD international conference
  on Knowledge discovery in data mining. ACM; 2005. p. 177--187.

\bibitem{wu2015emergent}
Wu Z, Menichetti G, Rahmede C, Bianconi G.
\newblock Emergent complex network geometry.
\newblock Scientific reports. 2015;5.

\bibitem{bhat2016densification}
Bhat U, Krapivsky P, Lambiotte R, Redner S.
\newblock Densification and structural transitions in networks that grow by
  node copying.
\newblock Physical Review E. 2016;94(6):062302.

\bibitem{bianconi2014triadic}
Bianconi G, Darst RK, Iacovacci J, Fortunato S.
\newblock Triadic closure as a basic generating mechanism of communities in
  complex networks.
\newblock Physical Review E. 2014;90(4):042806.

\bibitem{lambiotte2016structural}
Lambiotte R, Krapivsky P, Bhat U, Redner S.
\newblock Structural Transitions in Densifying Networks.
\newblock Physical review letters. 2016;117(21):218301.

\bibitem{snapnets}
Leskovec J, Krevl A. {SNAP Datasets}: {Stanford} Large Network Dataset
  Collection; 2014.
\newblock \url{http://snap.stanford.edu/data}.

\bibitem{rand2011dynamic}
Rand DG, Arbesman S, Christakis NA.
\newblock Dynamic social networks promote cooperation in experiments with
  humans.
\newblock Proceedings of the National Academy of Sciences.
  2011;108(48):19193--19198.

\bibitem{cuesta2015reputation}
Cuesta JA, Gracia-L{\'a}zaro C, Ferrer A, Moreno Y, S{\'a}nchez A.
\newblock Reputation drives cooperative behaviour and network formation in
  human groups.
\newblock Scientific reports. 2015;5:7843.

\bibitem{leskovec2012learning}
Leskovec J, Mcauley J.
\newblock Learning to discover social circles in ego networks.
\newblock Advances in neural information processing systems. 2012;25.

\bibitem{leskovec2007graph}
Leskovec J, Kleinberg J, Faloutsos C.
\newblock Graph evolution: Densification and shrinking diameters.
\newblock ACM transactions on Knowledge Discovery from Data (TKDD).
  2007;1(1):2--es.

\bibitem{raghavan2007near}
Raghavan UN, Albert R, Kumara S.
\newblock Near linear time algorithm to detect community structures in
  large-scale networks.
\newblock Physical review E. 2007;76(3):036106.

\bibitem{kirkpatrick1973percolation}
Kirkpatrick S.
\newblock Percolation and conduction.
\newblock Reviews of modern physics. 1973;45(4):574.

\bibitem{PhysRevLett.85.5468}
Callaway DS, Newman MEJ, Strogatz SH, Watts DJ.
\newblock Network Robustness and Fragility: Percolation on Random Graphs.
\newblock Phys Rev Lett. 2000;85:5468--5471.
\newblock doi:{10.1103/PhysRevLett.85.5468}.

\bibitem{masum2012reputation}
Masum H, Tovey M, Newmark C.
\newblock The reputation society: How online opinions are reshaping the offline
  world.
\newblock MIT Press; 2012.

\bibitem{easley2010networks}
Easley D, Kleinberg J.
\newblock Networks, crowds, and markets.
\newblock Cambridge Univ Press. 2010;6(1):1--6.

\bibitem{gilbert1959random}
Gilbert EN.
\newblock Random graphs.
\newblock The Annals of Mathematical Statistics. 1959;30(4):1141--1144.

\bibitem{serrano2006clustering}
Serrano M{\'A}, Bogun{\'a} M.
\newblock Clustering in complex networks. II. Percolation properties.
\newblock Physical Review E. 2006;74(5):056115.

\bibitem{hilbe2018indirect}
Hilbe C, Schmid L, Tkadlec J, Chatterjee K, Nowak MA.
\newblock Indirect reciprocity with private, noisy, and incomplete information.
\newblock Proceedings of the National Academy of Sciences.
  2018;115(48):12241--12246.

\bibitem{jordan2016third}
Jordan JJ, Hoffman M, Bloom P, Rand DG.
\newblock Third-party punishment as a costly signal of trustworthiness.
\newblock Nature. 2016;530(7591):473.

\bibitem{burt2004structural}
Burt RS.
\newblock Structural holes and good ideas.
\newblock American journal of sociology. 2004;110(2):349--399.

\bibitem{pentland2014social}
Pentland A.
\newblock Social physics: How good ideas spread-the lessons from a new science.
\newblock Penguin; 2014.

\end{thebibliography}
\end{document}


\begin{flushleft}
{\Large
\textbf\newline{Trust based attachment} 
}
\newline
\\
Julian Kates-Harbeck\textsuperscript{1},
Martin A. Nowak\textsuperscript{2,3*},
\\
\bigskip

\textbf{1} Department of Physics,
\\
\textbf{2} Department of Mathematics,
\\
\textbf{3} Department of Organismic and Evolutionary Biology, Harvard University, Cambridge MA~02138,~USA
\\
\bigskip

* martin\_nowak@harvard.edu 
\end{flushleft}

\renewcommand{\thetable}{\arabic{table}}   
\renewcommand{\thefigure}{\arabic{figure}}
\renewcommand{\figurename}{Extended Data Figure}
\renewcommand{\tablename}{Extended Data Table}
\setcounter{figure}{0}
\setcounter{table}{0}

\section*{Supplemental information}

\section*{SI Guide}
{\bf Supplemental Information}
This file contains supplemental discussions including sections with additional information on (i) the relationship with past work; (ii) details on methods and algorithms used; (ii) the calculation of the size of the expected audience; (iii) model assumptions and extensions; (iv) quantitative derivations around network generation; as well as (v) extended data figures.

\section*{Relationship with past work}

Our paper studies the relationship between indirect reciprocity and network structure. In the first part, we study how information spreads across an individual's neighborhood in order to quantify the expected size of the audience of a given interaction. In the second part, we use these insights to study trust based attachment.

\subsection*{Past work on indirect reciprocity}

The standard approach to indirect reciprocity assumes well mixed populations \cite{alexander1987biology,chalub2006evolution,wedekind2000cooperation,sigmund2010calculus,engelmann2009indirect,santos2018social,nowak1998evolution,brandt2005indirect,	nowak2005evolution,ohtsuki2004should,berger2011learning,radzvilavicius2018evolution} and does not take the effect of network structure on information flow into account.

\subsection*{Past work on information flow on networks}
Many papers have studied information flow and the spread of contagion on networks \cite{bakshy2012role,montanari2010spread,centola2018behavior,wu2008community,keeling2011insights,borge2013cascading,watts2002simple}, but these studies do not relate the spread of information to gossip and indirect reciprocity. Moreover, we study specifically the information spread as restricted to the neighborhood of any one given node. The work of Lind {\it et al.} \cite{PhysRevE.76.036117} does study this form of information spread. They consider the following model. For a given individual $i$ who is ``victim'' of the gossip, a neighbor $j$ spreads malicious gossip to all mutual neighbors of $i$ and $j$. In the next round, all those individuals spread the information to their mutual neighbors with the victim $i$, and so on. They average this process over all possible originators $j$ from $i$'s neighborhood. They then study the fraction of neighbors of $i$ that are eventually reached by the gossip ($f$), as well as the mean number of rounds to reach them ($\tau$), as a function of the degree of node $i$ on various real and artificial networks \cite{johansson2017gossip}. They find that for several network types, there exists an ideal degree for $i$ such that the spread of malicious gossip (as quantified by $f$) is minimized. They present various modifications of their original model in later sections. There exists an interesting relationship between their $f$ in the updated model introduced in the last 2 paragraphs of section IV, and our quantity $n_{ij}$:
$$f = \frac{1}{k_i}\left(1 + \sum_{j \in \mathcal{N}(i)}n_{ij}\right)\;,$$
where $k_i$ is the degree of node $i$ and $\mathcal{N}(i)$ is the set of neighbors of node $i$. The authors do not relate their analysis of gossip spread to indirect reciprocity and also do not consider the effect of gossip on network growth and attachment. By contrast, these are key components of our paper. Moreover, we consider information propagation for each neighbor of each focal individual (i.e. for all edges) on the network and analyze the local structural features affecting the spread. 

Shaw {\it et al.} \cite{shaw2011effect} study analytically and numerically the effect of random gossip on network structure, due to the ``bonding'' (i.e. the strengthening of ties) that occurs between two individuals who gossip about a third party ``victim'', and the weakening of ties between the victim and the gossipers. They find that if gossip spreads far, it helps build more cohesive clusters and strengthens triads, while it otherwise destroys triads. This influence of ``bonding'' is distinct from that of indirect reciprocity and thus forms a complimentary direction of study. 

\subsection*{Past work on indirect reciprocity on networks}
Several papers study models including information flow and link formation on networks in the context of repeated games \cite{pujol2005can,vega2006building,traag2011indirect,corten2014computational}. The authors focus on how an individual may update links based on the past behavior of other agents in the repeated game, and how the resulting networks co-evolve with individuals' strategies. Their finding that more cohesive, tightly clustered networks go hand in hand with cooperative behavior \cite{traag2011indirect,vega2006building} is in line with our result that such structures lead to higher values of $n_{ij}$. These papers do not study directly which network properties enhance or suppress the spread of gossip. 

\subsection*{Novelty of our approach}

\begin{itemize}

	\item{} We model the spread of gossip on social networks in the context of indirect reciprocity. This leads to an expected audience size $n_{ij}$ that is specific for any interaction between any pair of individuals $i$ and $j$ on the network.

	\item{} Our paper asks how network structure influences the power of indirect reciprocity. We explore the behavior of $n_{ij}$ for individual edges, and describe its dependence on local structural features of the neighborhood.

	\item{} We derive analytical approximations for the expected audience size $n_{ij}$ as a function of local network parameters.

	\item{} Due to indirect reciprocity, we consider the expected audience size as a measure of the incentives (the ``threat'' of reputation consequences or the temptation of reputational rewards) influencing  an actor's future behavior towards a recipient.

	\item{} Our approach models the effect of indirect reciprocity on the growth dynamics of social networks. We assume that individuals seek attachments where their neighbors have a high incentive to be cooperative due to the ability of the local network to conduct gossip. This results in the network growth mechanism of trust based attachment.

	\item{} In contrast to updating existing links given past experience, TBA represents a strategy for forming the first links in a novel network setting, thus complimenting the above body of work.

	\item{} Triadic closure is shown to be a good heuristic approximation for TBA.

\end{itemize}

\subsection*{Past work on models of network growth}

There is an extensive literature on models of network growth that can reproduce various aspects of the statistics \cite{orsini2015quantifying} of real world networks. Small world behavior can arise with a small number of random long-range connections in otherwise densely locally connected networks \cite{watts1998collective}. Models based on randomly connected groups of communities of various sizes and densities \cite{kolda2014scalable} can reproduce both clustering and degree statistics of target networks, although they require a global (i.e. not agent-based) algorithm to generate networks.  
Agent based models have shown scale free degree distributions to emerge from preferential attachment \cite{barabasi1999emergence}. A trade-off of popularity and similarity can accurately model additional properties of networks such as their hyperbolic geometry and high local clustering \cite{papadopoulos2012popularity}. These models however require global knowledge of the network (such as the degree of all nodes) for an agent to implement their attachment mechanisms. Triadic closure, i.e. forming a new link to a neighbor of a neighbor on the network, is an agent based mechanism common in realistic social settings \cite{bhat2014emergence,davidsen2002emergence,leskovec2005graphs}. Generative network models based on triadic closure with only few additional parameters can reproduce the characteristics of real social networks quantitatively \cite{wu2015emergent,bhat2016densification,bianconi2014triadic,lambiotte2016structural,leskovec2005graphs}.

\subsection*{Novelty of our approach} 
\begin{itemize}

	\item{} Our key innovation is to use indirect reciprocity to motivate growth.

	\item{} Our approach motivates trust based attachment by considering individuals' incentives in the presence of indirect reciprocity and reputation spread. Unlike all the above approaches, this provides a game theoretic justification for TBA as an attachment strategy.

	\item{} Our approach directly motivates the well known method of triadic closure (friend-of-friend attachment) by showing that it is a good heuristic approximation to TBA.

	\item{} Friend-of-friend attachment is a local, agent-based attachment algorithm which does not require global knowledge, whose only parameters are the total size of the network $N$ and its average degree $k$, and which is fast and easy to implement with a runtime complexity of only $O(N k)$.
\end{itemize}
Numerous authors have studied indirect reciprocity, the spread of gossip and information on social networks, as well as models of network growth and attachment. Here we present an analysis that joins indirect reciprocity and network growth and thus provides a missing link between those literatures.



\section*{Methods and Algorithms}

\subsection*{Calculating $n_{ij}$}
In general, computing $n_{ij}$ has exponential running time in the number of edges of the actor's neighborhood. We employ two methods for calculating $n_{ij}$ in this paper. For sufficiently small local neighborhoods (if the degree of an actor node is $\lesssim 8$), we calculate $n_{ij}$ exactly by enumerating all possible combinations of which edge is sampled in the neighborhood graph, calculating the audience size in that case, and summing all those values weighted by the probability of sampling those edges. For larger neighborhoods, we use Monte Carlo sampling to evaluate $n_{ij}$. For $N_{trials}$ repetitions, we sample the neighborhood graph (each edge is kept with probability $p$) and count the size of the audience. The values are summed up and the mean value over all $N_{trials}$ repetitions is reported. For all experiments we use $N_{trials} = 10^4$.

\subsection*{Testing the approximation for $n_{ij}$}
\begin{figure*}[p]
\noindent\makebox[\linewidth]{
  \includegraphics[width=1.0\linewidth]{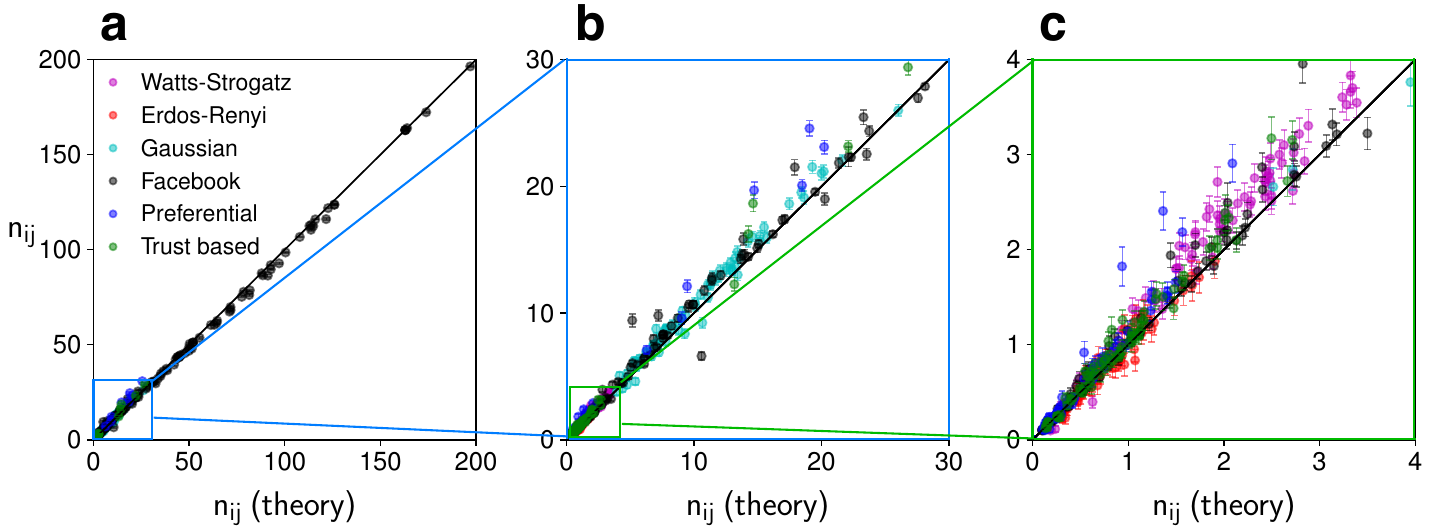}
}

\caption{{\bf Analytical approximation for calculating $n_{ij}$.} We show the numerically computed exact values of $n_{ij}$ vs. our theoretical approximation based on $k_i$, $c_i$ and $m_{ij}$ (see the section ``Random neighborhood approximation for $n_{ij}$'' in the \SI{} for details). Each point represents one randomly selected actor and recipient pair, and the pairs are sampled randomly from various artificial and real-world networks. The diagonal solid black line indicates equality. The different panels {\bf(a-c)} zoom into the same plot at various levels of detail. Error bars (s.d.) resulting from the numerical evaluation of $n_{ij}$ are given for each data point. The approximation is universally applicable across various networks sizes and structures. We use $2000$ samples for the Monte-Carlo computation of the exact values of $n_{ij}$. Network parameters --- ``Watts-Strogatz'': small world network \cite{watts1998collective} with $N = 200$ individuals and an average degree of $k = 20$; ``Erdos-Renyi'': erdos-renyi random graph with $N = 200$ and $k = 30$; ``Gaussian'': random 2D geometric random graph with a gaussian connection probability with $N = 200$ and $k = 40$, ``Facebook'': Facebook social network \cite{snapnets,leskovec2012learning} with $N =4039$ and $k = 44$; ``Preferential'': Network generated by preferential attachment \cite{barabasi1999emergence}; ``Trust based'': Network generated by trust based attachment (see main text) with $N = 200$ and $k = 20$; $p = 0.1$. We plot $80$ points for each graph, except the Facebook graph for which we plot $200$ points.
}
\label{edfig:simcompare}
\end{figure*}

We test numerically our assumptions and find that they hold well over a wide range of the relevant parameters, and all the different graph types that we tested. Using Monte-Carlo simulations, we find the exact value of $P$ in the context of the different reputation propagation models defined above (mutual neighbors only, local neighborhood model). These values are computed for several nodes sampled randomly from various different graph types, including Erdos-Renyi random graphs, Watts-Strogatz small world networks \cite{watts1998collective}, Random Gaussian Geometric Graphs (obtained by placing nodes uniformly at random in a square 2D region with periodic boundary conditions, and forming links with a Gaussian probability distribution as a function of Euclidean distance between node pairs, normalized such that the overall average degree on the network becomes some specified $k$), preferential attachment graphs \cite{barabasi1999emergence}, graphs generated by trust based attachment as described in this paper, and a subset of the social network graph from Facebook \cite{snapnets}. The results are shown in Extended Data \cref{edfig:simcompare}. The agreement between simulations and theory is universal, for all nodes and edges on all graphs studied. Our analytical and numerical results imply that for any node on any social network, the simple parameter combination of $k$, $c$, and $m_{ij}$, gives a very strong and precise prediction of the expected audience size $n_{ij}$.

\subsection*{Generating social networks}
Let us consider an individual in a population subject to an interaction structure. Given that the structure of social networks has a strong influence on other individuals' incentive to cooperate, how should our individual best attach themselves to other individuals? How can the individual exploit local population structure to increase their payoff \cite{rand2011dynamic,cuesta2015reputation}? The individual wishes to chose their neighbors such that these neighbors have a high incentive to cooperate, since this maximizes the payoff for the individual in question. Recall that $n_{ij}$ can be interpreted as a measure of the incentive of individual $i$ to cooperate with individual $j$ (note the directionality). Thus, it is a measure of \emph{trust} between $j$ and $i$. The higher this value, the more $j$ can trust that $i$ will cooperate (because it would suffer the reputational consequences otherwise). We consider the following specific algorithm inspired by this trust based attachment. A node $j$ enters an existing population encoded as a network. It then attaches $k/2$ edges into the population, one after the other, where the neighbors for these edges are chosen as follows. For every node $i$ on the graph to which $j$ is not connected yet, $n_{ij}$ is computed. One node is then sampled from these candidates proportional to its value of the trust measure $n_{ij}$, and an edge is formed to that new neighbor. The average clustering of the resulting graph can be controlled by the weighting of $n_{ij}$. The probability of picking a neighbor $i$ can in principle be a \emph{nonlinear} function of $n_{ij}$ that emerges from the importance of certain connections, how likely we are to keep them, or how often introductions are made; \cite{bhat2014emergence} for example is a threshold model in $n_{ij}$. We find using numerical simulations that the limiting value of the clustering coefficient is an increasing function of how strongly (i.e. compared to linear) the probability of attachment increases with $n_{ij}$.

For the first of the $k/2$ edges the node $j$ won't have any neighbors yet and thus all $n_{ij}$ will be zero. The first neighbor is therefore chosen at random. The results are not sensitive to this choice. The resulting networks show the same statistics if the first neighbor is chosen according to preferential attachment. To grow a population of $N$ individuals, we begin with a fully connected graph of $k+1$ individuals (to ensure that the average degree of the resulting network is exactly $k$) and then add one node at a time with $k/2$ edges each as described above until we have $N$ total individuals. The result of this attachment model are shown in {\bf Figure 3}.

This simple trust based attachment model generates graphs that show key characteristics of social networks \cite{wu2015emergent}: small world connectivity (i.e. path length that scales like $\sim log N$), high clustering coefficient that approaches a constant value as $N$ grows, and power-law scale free degree distributions.

Computing the exact value of $n_{ij}$ is computationally intensive and not necessarily a viable strategy for an individual with imperfect information about the surrounding social network. In the \SI{} section ``Size of the expected audience'', we provide an approximate analytical expression based on $n_{ij}$, $c_i$ and $k_i$. Moreover, we can find the following very simple approximation for $n_{ij}$:
$$n_{ij} \approx p m_{ij}\;.$$
This approximation is obtained in the limit where $p \to 0$ or where we only consider paths of gossip spread of length at most $1$. Either limit thus leads to the very simple rule that the probability of node $j$ picking a new edge to node $i$ is proportional to $m_{ij}$, the embeddedness of the edge $ij$.

This rule can now be interpreted (and implemented) as a very simple strategy by an individual $j$. For every new edge that node $j$ wants to add, do the following: {\J 1) pick a random neighbor; 2) pick one of their random neighbors; 3) Attach to them if there isn't already an edge. Repeat until $k/2$ edges have been formed.} We show below that this is equivalent to picking a neighbor proportional to $m_{ij}$. This is a very intuitive and easy to implement rule: simply attach yourself to a random friend of a friend. {\bf Figure 3} demonstrates that both trust based attachment with its full complexity, as well as the approximation based on attaching to a friend of a friend, result in the same macroscopic network properties qualitatively and quantitatively: high clustering, power law degree distributions, and small average path length. Moreover, it is simpler than preferential attachment in that it does not require global information. A naive implementation of friend-of-friend attachment for generating a network of average degree $k$ with $N$ individuals has $O(N k)$ runtime, compared with $O(N^2 k)$ for preferential attachment. Thus, this algorithm can be used to quickly generate realistic social network-like structures. 

\subsection*{Plotting social networks}
The data for the social networks is taken from the Google Plus, Facebook \cite{leskovec2012learning}, EU research email, and General Relativity collaboration networks \cite{leskovec2007graph} obtained from the Stanford Large Network Dataset Collection \cite{snapnets}. For the Google Plus network, we use the first $10^6$ edges. For the other networks, we use all edges.

Since visualization of such large networks is not practical, we focus on small subgraphs of these large networks when generating plots. In particular, we use the following approach to subsample large networks to desired values of $N$ and $k$ for plotting purposes. We first divide the larger network into communities of various sizes by running a standard community detection algorithm \cite{raghavan2007near}, and then find the community closest in $N$ to the desired value. We then remove random edges until the mean degree is as close as possible to $k$, and keep the largest connected component as our final sub-sampled network. For TBA, we first construct networks with the correct $N$ and twice the average degree as desired and then run the same sub-sampling procedure to obtain the desired $N$ and $k$. This provides a more fair comparison since all generated networks are now subject to sub-sampling, which can have effects such as introducing nodes with very few neighbors. This procedure is only used when we plot large example social networks and networks generated by TBA in {\bf Figure 4} as well as Extended Data \cref{fig:edfig_ni_networks,fig:top_c,fig:edfig6}. This procedure does not affect our computation of network statistics and metrics such as those shown in {\bf Figure 3} or Extended Data \cref{edfig:simcompare,fig:edfig_ni_stats,fig:graph_generation_extended}.

\section*{Size of the expected audience}
Consider the game as described in {\bf Figure 1} and the main text. We are interested in how $i$'s decision of whether to act altruistically depends on the $P_{jk}$, the probability that information reaches $k$, another neighbor of node $i$, from node $j$, as long as all nodes on the path from $j$ to $k$ are connected to node $i$. Information is said to successfully travel between two nodes if there is a path of edges between those two nodes, where each edge independently and at random transmits information with probability $p$. Thus, $P_{jk}$ is the edge percolation probability \cite{kirkpatrick1973percolation,PhysRevE.76.036117,PhysRevLett.85.5468} with parameter $p$ between node $j$ and a given neighbor $k \ne j$ of node $i$, on the graph defined by $\mathcal{N}(i)$, the subgraph defined by the neighborhood of node $i$. Intuitively, if the $P_{jk}$ are high, then neighbors are likely to find out about the good deed and reward the behavior. Thus $i$ will have a high incentive to perform the act. On the other hand, if the $P_{jk}$ are low, $i$ will not have  a large incentive to perform the good deed. The sum of $P_{jk}$ over all neighbors measures the total expected audience size of a given action of node $i$:
$$n_{ij} = \sum_{k \in \mathcal{N}(i), k \ne j} P_{jk}\;,$$
where $\mathcal{N}(i)$ denotes the neighborhood of node $i$.
It can also be viewed as how much neighbor $j$ can ``trust'' node $i$ to be cooperative. It depends crucially on the network structure and the resulting ability of the network to diffuse reputation in the neighborhood of that or edge. Moreover, more effective information transmission (higher $p$) also makes altruistic action more favorable. Reputation systems have recently been scaled up to provide security and trust for transactions in vast and otherwise anonymous online communities such as Ebay, AirBnb, or Tripadvisor \cite{masum2012reputation}.

\subsection*{Random neighborhood approximation for gossip propagation}\label{apdx:random_neighborhood}
Suppose we have an actor node $i$, its neighborhood $\mathcal{N}(i)$, and a given recipient $j \in \mathcal{N}(i)$. We are interested in finding the edge percolation probability $P_{jl}$ from $j$ to a given other neighbor $l \in \mathcal{N}(i),\; l \ne j$ on the subgraph defined by $\mathcal{N}(i)$. We assume that the following quantities are known: the degree $k_i$ of node $i$, the local clustering coefficient $c_i$ of node $i$ (defined as the probability that two randomly chosen neighbors of $i$ are connected to each other), as well as the embeddedness \cite{watts1998collective,easley2010networks} $m_{ij}$ of the edge $ij$, which is the number of mutual neighbors of nodes $i$ and $j$.

By definition, the size of the neighborhood subgraph is $k_i$ and the number of edges on that subgraph is $c_i*\frac{k_i (k_i-1)}{2}$. Moreover, the number of edges between the recipient (the originator of the percolation) and the rest of the neighborhood subgraph is by definition $n_{ij}$.

We now make the following key simplifying assumption: We assume that the edges are distributed randomly on the neighborhood subgraph (Extended Data \cref{fig:random_graph_approx}). In particular, we imagine that the recipient $j$ is connected to each of the rest of the $k_i - 1$ neighborhood nodes with a probability of $\frac{m_{ij}}{k_i-1}$. This results in $m_{ij}$ edges in expectation, which is the true number of such connections. Moreover, we assume that the rest of the $k_i - 1$ nodes are connected randomly (like an Erdos-Renyi random graph) among each other with probability 
$$c^* = \frac{c_i \frac{k_i (k_i-1)}{2} - m_{ij}}{\frac{(k_i-1)(k_i-2)}{2}}\;.$$
This again results in the true number of remaining edges $c_i \frac{k_i (k_i-1)}{2} - m_{ij}$ in expectation.

In the percolation process, each of these edges is now kept with probability $p$. Thus, in our approximation, the final subgraph after edges have been sampled for the percolation process now has random edges from the recipient to the rest of the neighborhood with probability
$$p_1 \equiv p \frac{m_{ij}}{k_i-1}\;,$$
and edges among the rest of the neighborhood with probability 
$$p_2 \equiv p \frac{c_i \frac{k_i (k_i-1)}{2} - m_{ij}}{\frac{(k_i-1)(k_i-2)}{2}}\;.$$
Without knowledge of $m_{ij}$, the equivalent best estimate would be to assume $p_1 = p_2 = c p$.

While these quantities are on average correct, the ``random graph'' assumption specifically assumes that there is no significant connectivity structure in this neighborhood (such as disconnected communities, etc.). Our assumption gets worse the less random the structure of the neighborhood is. This can be measured for example by the mean clustering coefficient on $g$. If it is that of a random graph (the lowest possible value), percolation is the most likely and our assumption is most accurate. The higher the clustering, the more our assumption overestimates $P$. Moreover, our assumption overestimates the percolation probability if the edges $m_{ij}$ are concentrated among a separate, tightly connected subset of the neighborhood instead of being randomly distributed among the whole neighborhood.

The final expression for the percolation probability is
\begin{equation}\label{eq:percolation}
P_{jl} = 1- \sum_{k'=1}^{k_i-1} A_{k'}(p_2)\binom{k_i-1}{k'}(1-p_2)^{k'(k_i-1-k')}\frac{k'}{k_i-1}(1-p_1)^{k'}
\end{equation}
where $A_j(x)$ is the reliability function \cite{gilbert1959random} that a graph of size $j$ is fully connected given that all edges exist independently and at random with probability $x$. This approximation of $P_{jl}$ is the same for all $l$. The final expectation value $n_{ij}$ is a sum over all $k_i - 1$ nodes on $g$:
$$n_{ij} = (k_i - 1) P_{jl}$$
with $P_{jl}$ as given above.

We now derive \cref{eq:percolation}. The recipient is connected to the each node in the rest of the neighborhood with probability $p_1$. The other neighbors among each other are connected with probability $p_2$. We thus have a situation with a random graph of $k_i - 1$ nodes, connected among each other with probability $p_2$ --- let us call this graph $g$ --- and a single outlying node (the recipient) connected to every node in $g$ with probability $p_1$. $P_{jl}$ is now the probability whether there is any path from the outlying node to a randomly chosen one of its neighbors. This situation is illustrated in Extended Data \cref{fig:random_graph_approx}.  

Let us find the complement probability -- that a randomly chosen node --- call it $l$ --- on $g$ is not connected to the recipient (i.e. there is no path between them). In particular, consider a connected subgraph of $g$ of size exactly $k'$. We will find the probability that node $l$ is in this subset of size exactly $k'$, and that the recipient is \emph{not} connected to it.

Pick a subset of $k'$ nodes. The probability of this subset being connected is $A_{k'}(p_2)$, and there are $\binom{k_i-1}{k'}$ such subsets. The probability that this subset is disconnected from all other $k_i - 1 - k'$ nodes on $g$ is $(1-p_2)^{k'(k_i - 1 -k')}$ (i.e. the connected subset has size exactly $k'$). With probability $\frac{k'}{k_i - 1}$, the node $l$ is part of this subset. We need to make sure that the recipient is disconnected from all nodes in the subset -- this is given by probabilty $(1-p_1)^{k'}$. Thus, a single term (for a given $k'$) in the sum in equation \ref{eq:percolation} is the probability that node $l$ lies in a connected subset of size exactly $k'$, and that the recipient is not connected to this subset. By summing over all possible sizes $k'$, we have listed all disjoint ways in which the recipient can be disconnected from node $l$. Thus, this is by definition $1-P_{jl}$, giving equation \ref{eq:percolation}.

In the case where $p_1 = p_2 = cp$, this reduces to the ``two-terminal reliability function'' $T_{k_i-1}(c_i p)$ \cite{gilbert1959random}, which measures the probability that two randomly chosen nodes on a random graph of size $k_i -1$ with edge probability $c_i p$ are connected.

\subsection*{Limiting cases of $P_{ij}$}\label{apdx:P_ij_limits}
Consider the limit as $p \to 0$. In this limit, only paths of length $1$ will contribute to the flow of information, since all other paths only exist with probability $O(p^2)$. By definition, there are $m_{ij}$ edges (paths of length $1$) from the recipient to the rest of the neighborhood $\mathcal{N}(i)$ of the actor. Each such edge independently carries information with probability $p$. Thus we have
$$n_{ij} = p m_{ij} + O(p^2)\;.$$
This expression becomes exact as $p \to 0$ and makes the strong dependence of $n_{ij}$ on the embeddedness $m_{ij}$ explicit. The dependence of $n_{ij}$ on features of the neighborhood structure beyond the embeddedness is thus a ``higher order'' effect and becomes more important for larger $p$.

\section*{Model assumptions and extensions}\label{sec:model_extensions}
As in any model, our approach involves several simplifying assumptions which we discuss here. Overall however our goal is not to model the full complexities of indirect reciprocity, but rather to isolate the impact of social network structure in particular on reputation diffusion and the consequences thereof on cooperation.

\subsection*{Global spread}
We have assumed that reputation can only travel across nodes that are neighbors of the central node. One could also consider a model where reputation can travel across any path on the entire network. We consider a probability $p_1$ of transmitting gossip on the local neighborhood of the actor, and a probability $p_2$ on any other edge of the graph. This model reduces to our original ``local'' model with $p_1 = p$ as $p_2 \to 0$. This kind of global information spread would entail that people gossip about third parties that they don't actually know (of). One example of this is two people talking about a celebrity.

In this case, the global structure of the network becomes relevant. An interesting tradeoff arises: there is now an exponentially growing number of possible paths for the reputation to flow, however, longer paths are exponentially less likely to exist. We thus have the percolation probability on the overall graph between two nodes becoming relevant. We find numerically that in this model the overall behavior of the percolation probability now depends on global parameters of the graph such as the overall number of nodes, mean degree and global clustering. A way to approximate the results would be to use insights from the global percolation properties of clustered networks \cite{serrano2006clustering}. 

We study this model numerically in Extended Data \cref{fig:global_percolation}. We show example pairs of actors and recipients from various different networks, and the resulting values of $n_{ij}$ with a given value of $p_1$ and as a function of $p_2$. As expected, the global percolation model reduces to the known local solution as $p_2 \to 0$. For finite values, the value of $n_{ij}$ increases. The probability of information traveling from the recipient to another neighbor of the actor via edges outside of the actor's neighborhood depends on the percolation properties of the overall network. 

We show the global edge percolation threshold $p_c$ on the overall network (determined numerically), defined as the probability of keeping a given edge on the overall network for which the probability that two randomly chosen nodes remain connected via some path is $0.5$. The connection probability of two randomly chosen nodes rapidly becomes $0$ below this threshold, and $1$ above this threshold. The transition becomes sharper as networks increase in size \cite{kirkpatrick1973percolation}. The actor is by definition a mutual connection between the recipient and any other neighbor of the actor. This introduces a bias that distinguishes the recipient and other neighbors of the actor from two random nodes. Nevertheless, we find that the boost in $n_{ij}$ from the global percolation is strongly related to the global percolation probability as a function of $p_2$. In particular, we can in general distinguish the following regimes. For values of $p_2 < p_c$, the global value of $n_{ij}$ is nearly the same as for the local case with $p_2 = 0$. For values of $p_2 > p_c$, $n_{ij}$ rapidly approaches its maximum value $k-1$.
The only exception are cases where the network becomes disconnected once the actor node is removed, since the actor node and its edges are never counted in the transmission of gossip. In those cases, $n_{ij}$ approaches some value $< (k-1)$ for $p_2 > p_c$. Finally, there is an intermediate regime $p_2 \approx p_c$. Here the value of $n_{ij}$ rapidly rises from its local value to the maximum possible value $k-1$ with increasing $p_2$. The exact functional dependence depends on the full global network. Further aspects of possible models for the spread of gossip and their relation to global network properties are also explored in \cite{PhysRevE.76.036117}.

Overall, the question of who ``knows of'' another individual could also be tackled by considering directed networks in future work. Person $A$ might know (of) person $B$ but not vice versa, especially for the classic example of celebrities. Moreover the interaction graph might not be the same as the graph of who knows whom. Interactions and communication strength might be weighted. An individual with hundreds of connections may not be able to communicate with each neighbor to the same degree as an individual with only a few. While adding more parameters and complexity, a more complete model could thus consider directed and weighted graphs, as well as separate graphs for game interaction and reputation diffusion.

\subsection*{Global cooperation}
In the main text, we assume that the good deed of the actor is directed towards a particular recipient individual, who is then the originator of gossip. We may also consider the case of global cooperation, where the good deed is not directed towards anyone in particular, such as helping a stranger in need, performing public service, or making a donation. In this case, we consider the same model with the modification that every neighbor can independently and at random ``observe'' the good deed of the actor with probability $q$. In this case, all neighbors that observed the act serve as originators of gossip, which travels in the same way as in our original model with parameter $p$ (there now of course can be several originators). For any given node $i$ on the graph, the parameters $q$ and $p$ then define an expected audience size $n_i$ for this global cooperation. This quantity measures the incentive for any individual on the graph to be globally cooperative.

If we again approximate the neighborhood of $i$ as a random graph with $k_i$ individuals and a connection probability of $c_i$, we can use similar reasoning as in the section ``Random neighborhood approximation for gossip propagation''above to derive an approximate expression for $n_i$. In particular, let $\tilde{P}_{i}$ denote the percolation probability of the information reaching a randomly chosen neighbor $j$ of individual $i$ (the information needs to first travel to any neighbor of $i$ via observation, and then possibly via gossip to the neighbor $j$). Then with the random neighborhood assumption we have the same exact situation as in the derivation of equation \ref{eq:percolation} above, with the replacements $p_1 \to q$, $p_2 \to c p$, and $k_i - 1 \to k_i$. We thus obtain
\begin{equation}\label{eq:percolation_global}
\tilde{P}_{i} = 1- \sum_{k'=1}^{k_i} A_{k'}(c p)\binom{k_i}{k'}(1-c p)^{k'(k_i-k')}\frac{k'}{k_i}(1-q)^{k'}\;.
\end{equation}
Since this is the probability of the information reaching any particular neighbor, the expected number of neighbors reached by the gossip is then simply
$$n_i = k_i \tilde{P}_{i}\;.$$

In Extended Data \cref{fig:edfig_ni_stats} we show that this analytical approximation matches the real values of $n_i$ well for a wide variety of real and artificial networks. Moreover, we find empirically that the single variable $m_i$, the embeddedness of node $i$, is a very good predictor of the percolation probability $\tilde{P}_{i}$ and thus of the expected audience size. We define the embeddedness of a node as the average embeddedness of all edges attached to that node:
$$m_i = \frac{1}{k_i} \sum_{j \in \mathcal{N}(i)} m_{ij}\;.$$
We also show below the identity $m_i = (k_i - 1) c_i$. Universally, knowing only the degree and clustering (and thus the embeddedness) of a node allows us to make accurate predictions about its expected audience size $n_i$. 

In Extended Data \cref{fig:edfig_ni_neighborhoods} we illustrate the dependence of $n_i$ on the neighborhood structure of $i$. Neighborhoods that are highly connected generate the highest expected audiences, while neighborhoods consisting of several isolated individuals create low audiences. Finally, in Extended Data \cref{fig:edfig_ni_networks}, we show real world networks, where the nodes are colored by their values of $n_i$. Individuals with many neighbors that are all densely connected among each other have the largest values of $n_i$. Moreover, for those nodes, the power of indirect reciprocity for incentivizing global cooperation grows strongly with the probability $p$ of gossip transmission.

\paragraph{Relating the embeddedness $m_i$ to the degree and clustering}
Consider a node $i$ with local clustering coefficient $c_i$, degree $k_i$ and its neighborhood $\mathcal{N}(i)$. Then
$$m_i = \frac{1}{|\mathcal{N}(i)|} \sum_{j \in \mathcal{N}(i)} m_{i j} = \frac{1}{k_i} \sum_{j \in \mathcal{N}(i)} m_{i j}$$
now let $e_{j l} = 1$ if an edge exists between node $j$ and node $l$ and $e_{jl} = 0$ otherwise ($e_{jj} = 0$). Then $m_{ij} = \sum_{l \in \mathcal{N}(i)}  e_{j l}$
$$m_i = \frac{1}{k_i} \sum_{j \in \mathcal{N}(i)} \sum_{l \in \mathcal{N}(i)}  e_{j l} = \frac{1}{k_i} \sum_{j,l \in \mathcal{N}(i), j\ne l} e_{j l} \equiv \frac{1}{k_i} k_i(k_i-1) c_i = (k_i - 1) c_i$$
In the third equality we have used the fact that the expression $\frac{1}{k_i} \sum_{j,l \in \mathcal{N}(i), j\ne l} e_{j l} $ is simply twice (due to double counting $jl$ and $lj$) the number of edges that exist between all neighbors of $i$. There are $\frac{1}{2}k_i(k_i-1)$ possible edges, and the fraction that do exist is given by definition by the local clustering coefficient $c_i$. Thus, the number that do exist is $\frac{c_i}{2}k_i(k_i-1)$. Multiplying with the factor of $2$ due to double counting we obtain the above result. 

\subsection*{Symmetry of $n_{ij}$}
The definition of $n_{ij}$ is fundamentally asymmetric in $i$ and $j$. However, as shown in the main text, the main first-order dependence of $n_{ij}$ for small $p$ is on the number of mutual neighbors between $i$ and $j$, which is a symmetric quantity. Therefore, while network growth could arguably depend on a symmetric process of link formation that requires ``consent'' from both parties, we do not in this work consider alternative symmetric definitions of $n_{ij}$ for the numerical experiments for generating networks, as any symmetric definition capturing information flow on the local friendship network would lead to qualitatively similar behavior. 

\subsection*{Subtleties of moral judgment}
We assume that every defection carries universally negative reputation, while a cooperation always carries positive reputation. Here we do not consider further ``moral'' subtleties that arise in judging the reputation value of actions \cite{santos2018social,hilbe2018indirect}, such as whether defecting on a known defector is good or bad \cite{ohtsuki2004should,nowak2005evolution}. We also do not consider the ability of actors to make strategic decisions of which information to pass on. While these questions are important, our goal is to isolate the effect of social network structure on the spread of reputation. Thus, we choose the simplest model that captures these effects while deliberately not attempting to include the impact of moral judgment. 

\subsection*{Transmission of negative information}
In our model we only consider the transmission of distinctly positive information. One might also imagine a model where reputation spread can transmit distinctly negative information that would cause a greater punishment \cite{jordan2016third} or smaller payoff for that node in the future if neighbors find out. Either way, the size of the expected audience determines the strength to be cooperative or to not act negatively. 

\subsection*{Other connection incentives}
When considering trust based attachment, we have also assumed that considerations of future costs and benefits capture all relevant tradeoffs in forming, keeping, and breaking ties to other nodes. Naturally, there might be other reasons such as bridging structural holes, popularity, or similarity \cite{burt2004structural,pentland2014social,barabasi1999emergence,papadopoulos2012popularity} that can modify an individual's incentives for attachment on real social networks. Again, our model aims to isolate the effect of reputation and thus deliberately ignores these other incentives.

\section*{Network generation}
\subsection*{Picking proportional to $m_{ij}$ and picking random neighbor's neighbor is equivalent}\label{apdx:graph_generation}

Assume we have a node $j$ and pick a random neighbor's neighbor $i$. Consider a given such node $i$. There are then $m_{ij}$ possible paths of arriving at that node $i$ (one for every common neighbor of $i$ and $j$). Each of these paths has a specific probability of being chosen. The probability of ending up at $i$ is the sum of these ``weights'' of the various paths. Each path's weight (i.e. its probability of being chosen) depends on the degree of the common neighbor. However, if we assume that degree correlations on the graph aren't too strong, then all paths have in expectation the same weight, independently of the degree of $j$. Thus, in expectation, the probability of being picked is simply proportional to the number of such paths, $n_{ij}$.

We show this quantitatively. What is the probability of picking node $i$ via an intermediary node $l$? We will assume that the nearest neighbor degree distribution on our graph is approximately $P_{nn}(k) = \frac{k_l P(k_l)}{\bar{k}}$. Thus, this distribution is the same as the overall degree distribution, but each degree is upweighted by a factor of $k$. This is the default for graphs without degree correlations and emerges from the intuition that higher degree nodes have $k$ times more edges that one could be attached to. We have checked numerically that the graphs generated according to our model do not have strong degree correlations. $\bar{k}$ is the average degree on the graph. We will condition on the degree $k_l$ of node $l$ and sum over all possibilities. Recall that node $l$ is picked as a random neighbor of node $j$, and then $i$ is picked as a random neighbor of $l$.
\begin{eqnarray*}
P(\text{pick } i \text{ via } l) &=& \sum_{k_l} P(\text{pick } i \text{ via } l, k_l) \\
&=& \sum_{k_l} P(\text{pick } i \text{ via } l| k_l) P_{nn}(k_l)  \\
&=& \sum_{k_l} \frac{1}{k_j} \frac{1}{k_l} \frac{k_l P(k_l)}{\bar{k}} \\
&=&  \frac{1}{k_j \bar{k}} \sum_{k_l} P(k_l) \\
&=&  \frac{1}{k_j \bar{k}}\\
\end{eqnarray*}
In the third line, the first term is the probability of picking a given neighbor of $j$ (of which there are $k_j$). The second is the probability of picking a given neighbor of $l$. $P(k_l)$ is the degree distribution evaluated at $k_l$. Thus, each of the $m_{ij}$ paths to node $i$ has equal probability (independently of $i$), so the total weight is proportional to $m_{ij}$. Thus, picking random neighbor's neighbors is equivalent to picking a given node with probability proportional to $m_{ij}$. 

Due to its symmetry, the method also leads to two-way trusting relationships, since the methods forms new edges $ij$ proportional to $m_{ij}$, which is the number of mutual neighbors and thus symmetric, ensuring trustworthiness in both ways.

\subsection*{Why friend-of-friend generates a power law degree distribution}
Consider the distribution of degrees among the nodes that are being picked. Regardless of the degree of the intermediate node $l$, the degree of $i$ is drawn also from the nearest neighbor distribution $P_{nn}(k_i) = \frac{k_k P(k_i)}{\bar{k}}$. Compare this to the Barabasi-Albert model, where we pick a potential neighbor as a random node from the overall graph (which has degree distribution $P(k)$), with probability proportional to its degree. Thus, picking a node with degree $k$ has probability proportional to $P(k) k$. This is exactly the nearest neighbor degree distribution. Thus, the degree of the new neighbor in friend-of-friend attachment is drawn from the same distribution as in the BA model.  We therefore expect to see the same scale free degree distribution overall, while the friend-of-a-friend strategy additionally gives rise to high clustering and embeddedness. The power law emerges naturally from the bias of the nearest neighbor degree distribution and does not require assumptions of popularity or preferential attachment to popular nodes.

\subsection*{Departure of TBA and friend-of-friend for large $p$}
As shown above, the statistical equivalence between TBA and friend-of-friend attachment is only exact in the limit as $p \to 0$. In practice, we find (see Extended Data \cref{fig:graph_generation_extended}) that the two methods generate networks with very similar statistics up to moderate values of $p \lesssim 0.25$. For larger values, TBA generates degree distributions with departures from the power law form in the tail of the distribution (even more hubs are generated, which also results in even slower growth of the mean path length). The fact that real life networks do follow power law distributions may suggest that they are generated not directly by TBA, but rather by a heuristic of TBA like friend-of-friend attachment.

\section*{Extended data figures}

\begin{figure*}[p]
\noindent\makebox[\linewidth]{
  \includegraphics[width=1.0\linewidth]{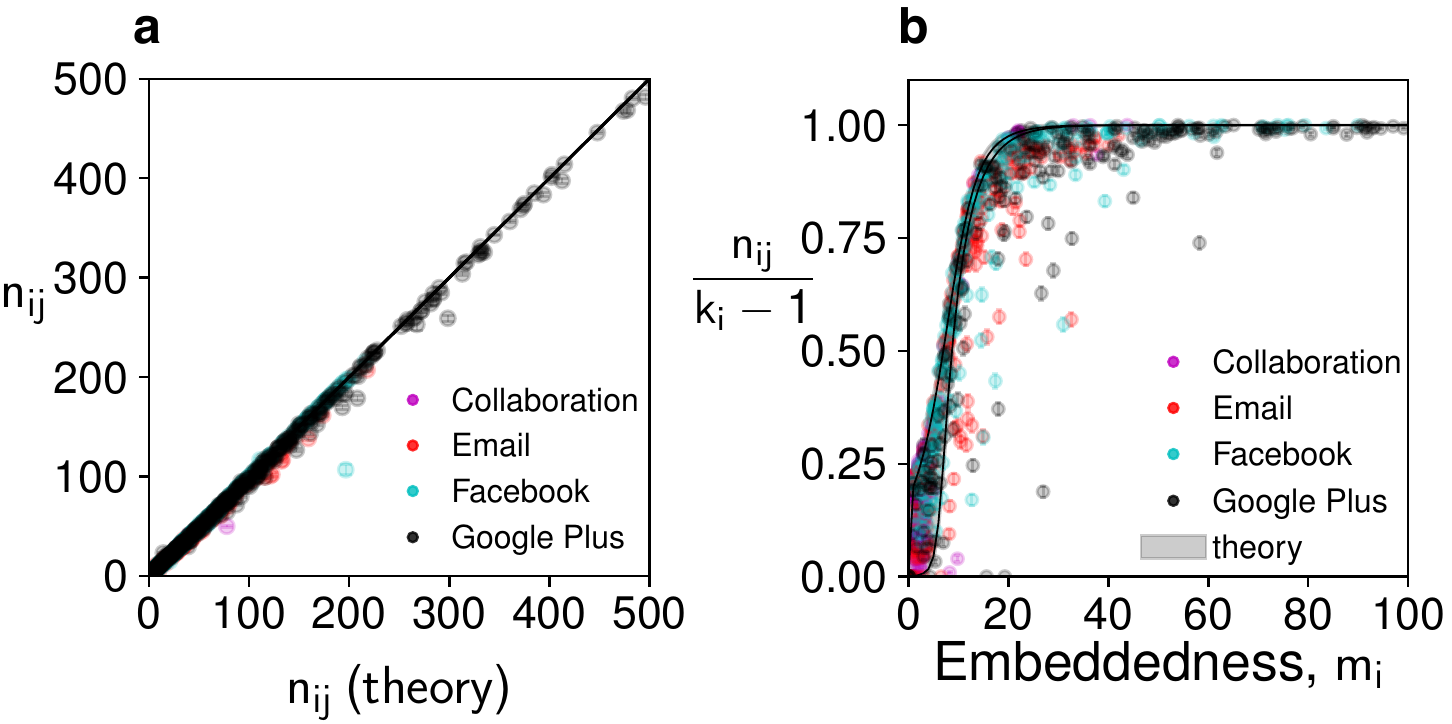}
}

\caption{{\bf Analytical approximation for calculating $n_{ij}$ on real networks.}
{\bf a,} Just as for the artificial networks in {Extended Data} \cref{edfig:simcompare}, our full analytical prediction based on $k_i$, $c_i$, and $m_{ij}$ (the embeddedness of the edge $ij$) agrees well with observed values on real networks (\cite{snapnets,leskovec2007graph,leskovec2012learning}, see ``Size of the expected audience'' in the \SI{} for details). {\bf b,} As a function of the embeddedness of the actor, the values of $n_{ij}$ approximately collapse onto {\J a narrow range, close to a univariate dependence}. We also show how this near collapse as a function of $m_i$ alone is also predicted by our analytical approximation for $n_{ij}$. In grey, we show the ranges of $n_{ij}$ predicted by our theory constrained by $m_i$ only, obtained by sweeping over possible values of $k_i \in (m_i + 1,N-1)$ for each $m_i$, fixing the resulting value of $c_i$ using the identity $m_i = (k_i-1) c_i$, and using our analytical prediction without using knowledge of $m_{ij}$ (see the \SI{} section ``Random neighborhood approximation for gossip propagation'' for details). In both plots, error bars are shown for each point, which are in most cases smaller than the plotted points. The points are for the same edges and the same graphs as shown in Fig. 2 in the main text. Per graph, $500$ randomly selected values of $n_{ij}$ are shown. Parameters: $p = 0.2$.
}
\label{edfig:simcompare_real}
\end{figure*}

\begin{figure*}[p]
\noindent\makebox[\linewidth]{
  \includegraphics[width=1.0\linewidth]{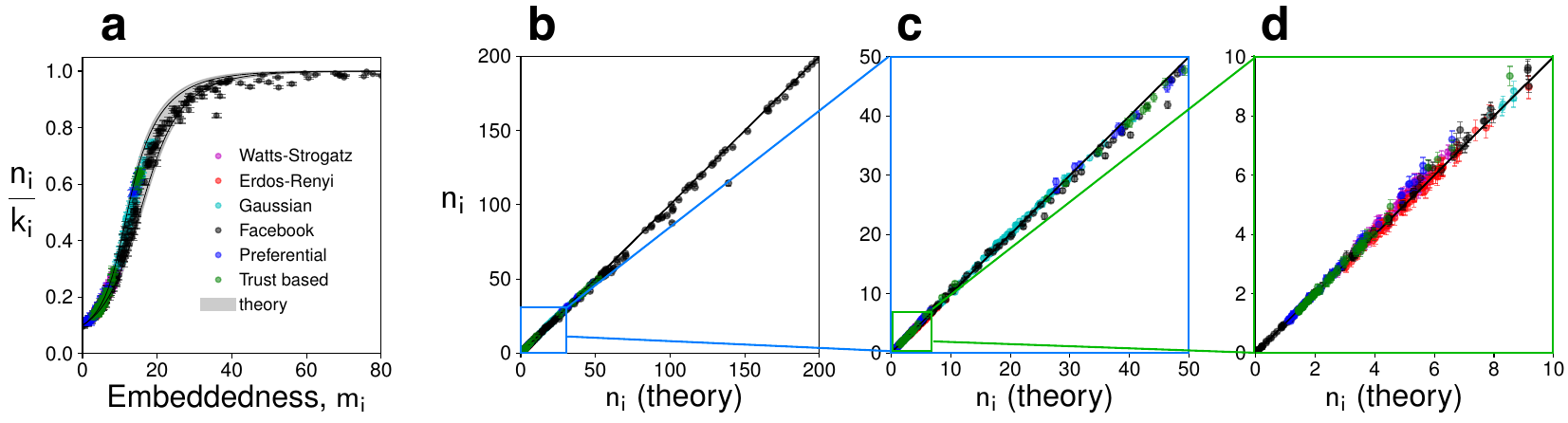}
}

\caption{{\bf Analytical approximation for calculating $n_{i}$, the expected audience size for global cooperation.} We show the numerically computed exact values of $n_{i}$ vs. our theoretical approximation based on $k_i$ and $c_i$ (see the section ``Global cooperation'' in the \SI{} for details). Each point represents one randomly selected actor $i$ sampled randomly from various artificial and real-world networks (the same networks as in {Extended Data} \cref{edfig:simcompare_real}). {\bf(a)} The dependence of the relative audience size $\frac{n_i}{k_i}$ is empirically found to be very well approximated by the single parameter $m_i$, the embeddedness of node $i$, which can be written as $m_i = (k_i - 1) c_i$ (see ``Global cooperation'' in the \SI{}). Both the theoretically predicted values (grey region) as well as the actual values of $n_i$ nearly collapse onto a line as a function of $m_i$. The diagonal solid black line indicates equality. The different panels {\bf(b-d)} illustrate the agreement between the analytical approximation and actual values of $n_i$. They zoom into the same plot at various levels of detail. Error bars (s.d.) resulting from the numerical evaluation of $n_{i}$ are given for each data point. The approximation is universally applicable across various networks sizes and structures. We use $2000$ samples for the Monte-Carlo computation of the exact values of $n_{i}$. Network and plotting parameters are the same as in Extended Data \cref{edfig:simcompare}. 
}
\label{fig:edfig_ni_stats}
\end{figure*}

\begin{figure*}[p]
\noindent\makebox[\linewidth]{
  \includegraphics[width=1.0\linewidth]{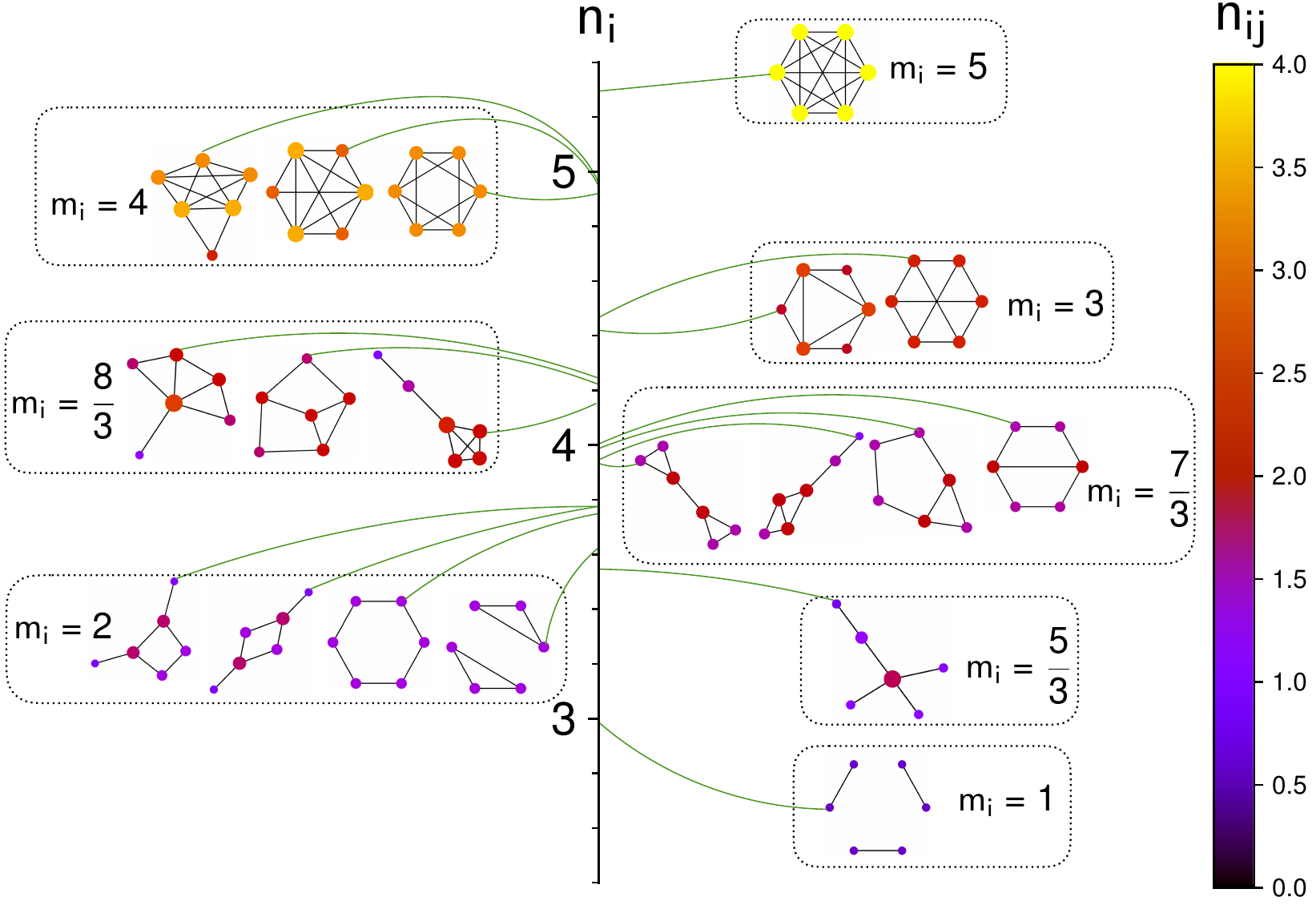}
}

\caption{{\bf Dependence of $n_i$ on neighborhood structure for various neighborhoods of size $6$}. Each graph represents the neighborhood structure of an actor node $i$. The actor itself is not shown for clarity. Each node is colored by its value of $n_{ij}$ that the actor has with respect to that node. The entire neighborhood is connected with a thin green line to the value of $n_i$ that the actor has for that neighborhood. The neighborhoods are grouped by the embeddedness $m_i$, which is highly predictive of the expected audience size $n_i$. Densely connected neighborhoods create a high expected audience for global cooperation $n_i$, and centrally connected individuals in the neighborhood enjoy a high expected audience size $n_{ij}$. Sparsely connected and disjoint neighborhoods have low values of $n_{i}$ and isolated neighbors have low values of $n_{ij}$. Parameters: $p = 0.4$.}
\label{fig:edfig_ni_neighborhoods}
\end{figure*}

\begin{figure*}[p]
\noindent\makebox[\linewidth]{
  \includegraphics[width=1.0\linewidth]{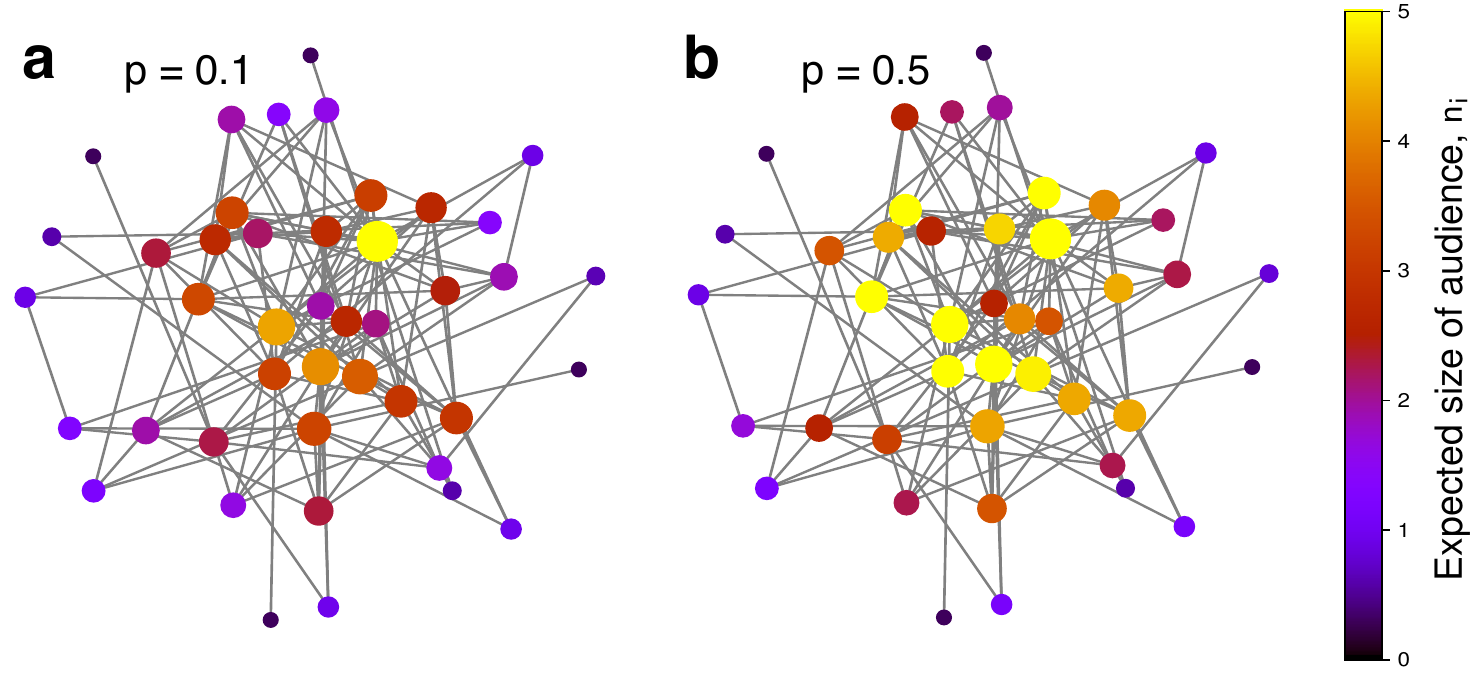}
}

\caption{{\bf Illustration of $n_i$ for a real social network}. The network is a subsample of the ``Facebook'' social network \cite{snapnets,leskovec2012learning}. Nodes are colored by the expected audience size for global cooperation, $n_{i}$. The two panels show the same network, but for low {\bf(a)} and high {\bf(b)} values of $p$. This value is a measure of the incentive for acts of global cooperation (see the section ``Global cooperation'' in the \SI{}). Nodes that are embedded in densely connected, large neighborhoods have high expected audience sizes. Moreover, their incentives grow strongly if the strength of gossip propagation $p$ increases. By contrast, isolated or weakly connected nodes have low values of $n_i$ and increased information flow does not increase their incentives much. Parameters: $N = 40$, $k = 6$, $p = 0.3$.}
\label{fig:edfig_ni_networks}
\end{figure*}

\begin{figure*}[p]
\noindent\makebox[\linewidth]{
  \includegraphics[width=0.99\linewidth]{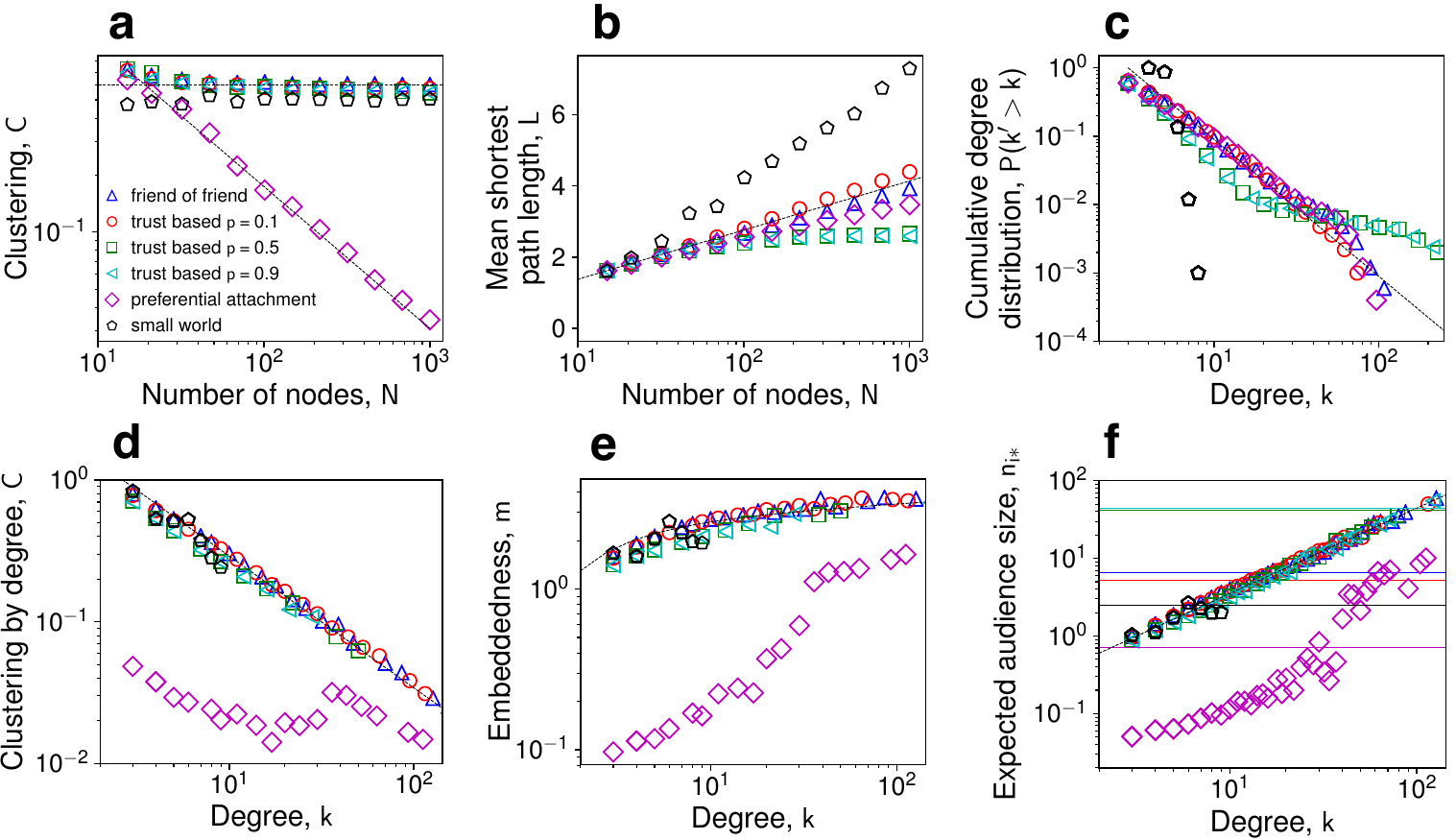}
  }

\caption{
{\bf Further characterization of graphs generated by trust based attachment.} {\bf (a-c)} The same statistics as shown in {\bf Figure 3}, but also for TBA graphs with $p \in \{0.5,0.9\}$, as well as small world and preferential attachment networks. {\J TBA with high values of $p \gtrsim 0.2$ shows departures from the friend-of-friend behavior. The degree distribution includes more hubs with even larger degree, and the resulting mean shortest path lengths grow sub-logarithmically with $N$. Clustering is nearly unchanged.}. Clustering is high and decreases as a power law with slope $\sim -0.92 > -1$ (dashed line) for trust based attachment {\bf(d)}. Clustering is consistently low and decreasing with $k$ for preferential attachment. {\bf (e)} The average embeddedness $n = c (k-1)$ --- the key indicator for trustworthiness --- scales as $c(k) (k-1) \sim k^{-0.92} (k-1)$ (dashed line) and thus grows for higher values of $k$. {\bf (f)} The mean outgoing expected audience size $n_{i*}$ ($p = 0.5$) scales as a power law with slope $\sim 1.1$ (dashed line) as a function of $k$ for the TBA and friend of friend networks. The small world network has a near uniform distribution and the preferential attachment network has significantly lower values of $n_{i*}$. The mean value (averaged over all directed edges) for each network is shown as a horizontal line of the equivalent color. The TBA and friend of friend networks lead to the highest average values of $n_{i*}$. {\J We find that while these results qualitatively are independent of $p$, the slope of the power law and the global mean value increase with $p$}. All networks have final size $N = 1000$ and $k = 6$.}
\label{fig:graph_generation_extended}
\end{figure*}

\begin{figure*}[p]
\noindent\makebox[\linewidth]{
  \includegraphics[width=1.0\linewidth]{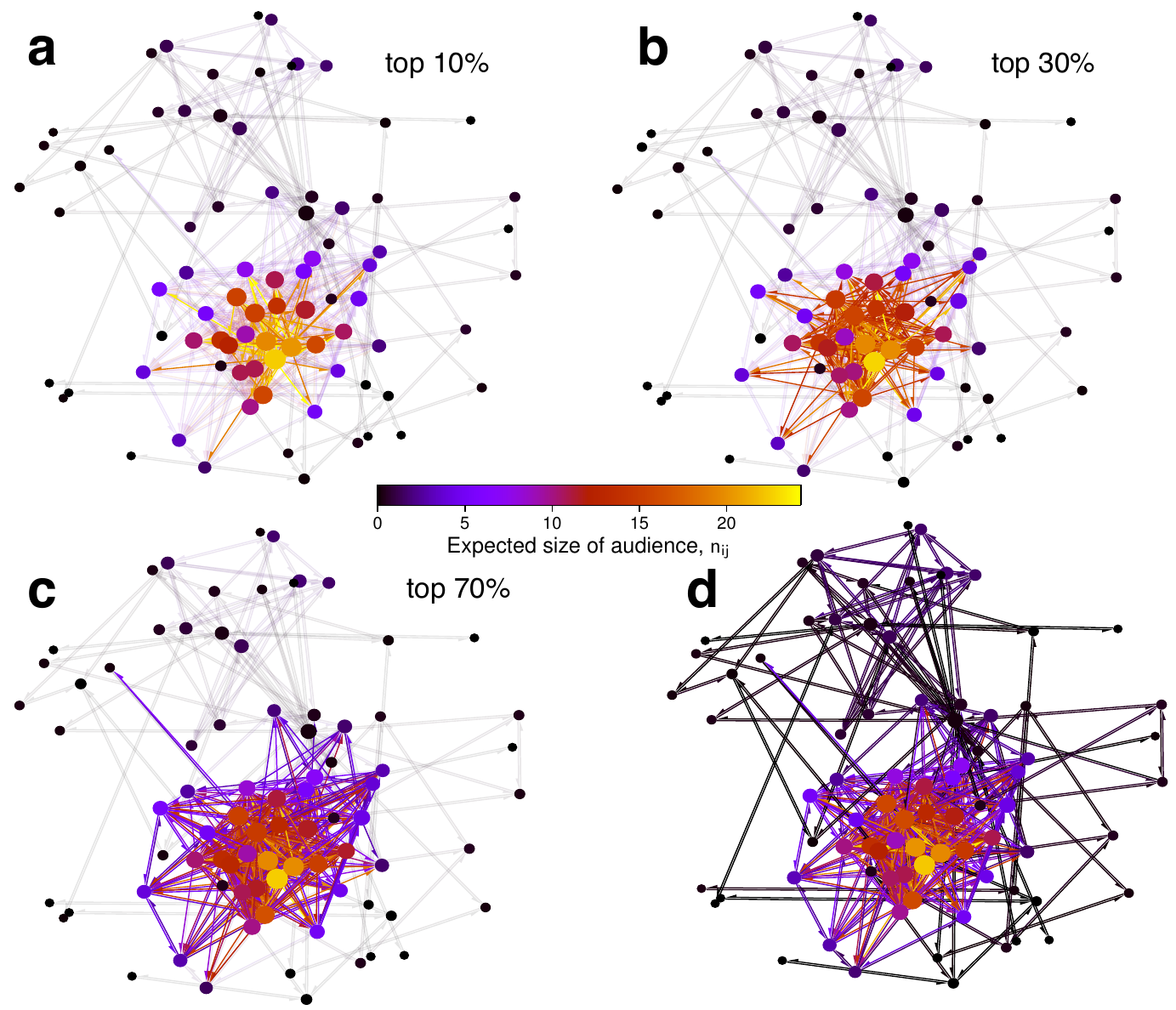}
}
\caption{
{\bf Illustration of the hierarchy of $n_{ij}$ values on a real social network}. The network is a subsample of the ``Facebook'' social network (\cite{snapnets,leskovec2012learning} see ``Plotting social networks'' in the \SI{} for details). Edges and nodes are colored as in {\bf Figure 4}. The panels {\bf(a-d)} show progressively larger fractions of the highest $n_{ij}$ values on the network. The highest values occur in densely connected communities, while the lower values occur in individuals that are less embedded and have fewer neighbors. $N = 50$, $k = 6$, $p = 0.4$.}
\label{fig:top_c}
\end{figure*}

\begin{figure*}[p]
\noindent\makebox[\linewidth]{
  \includegraphics[width=1.0\linewidth]{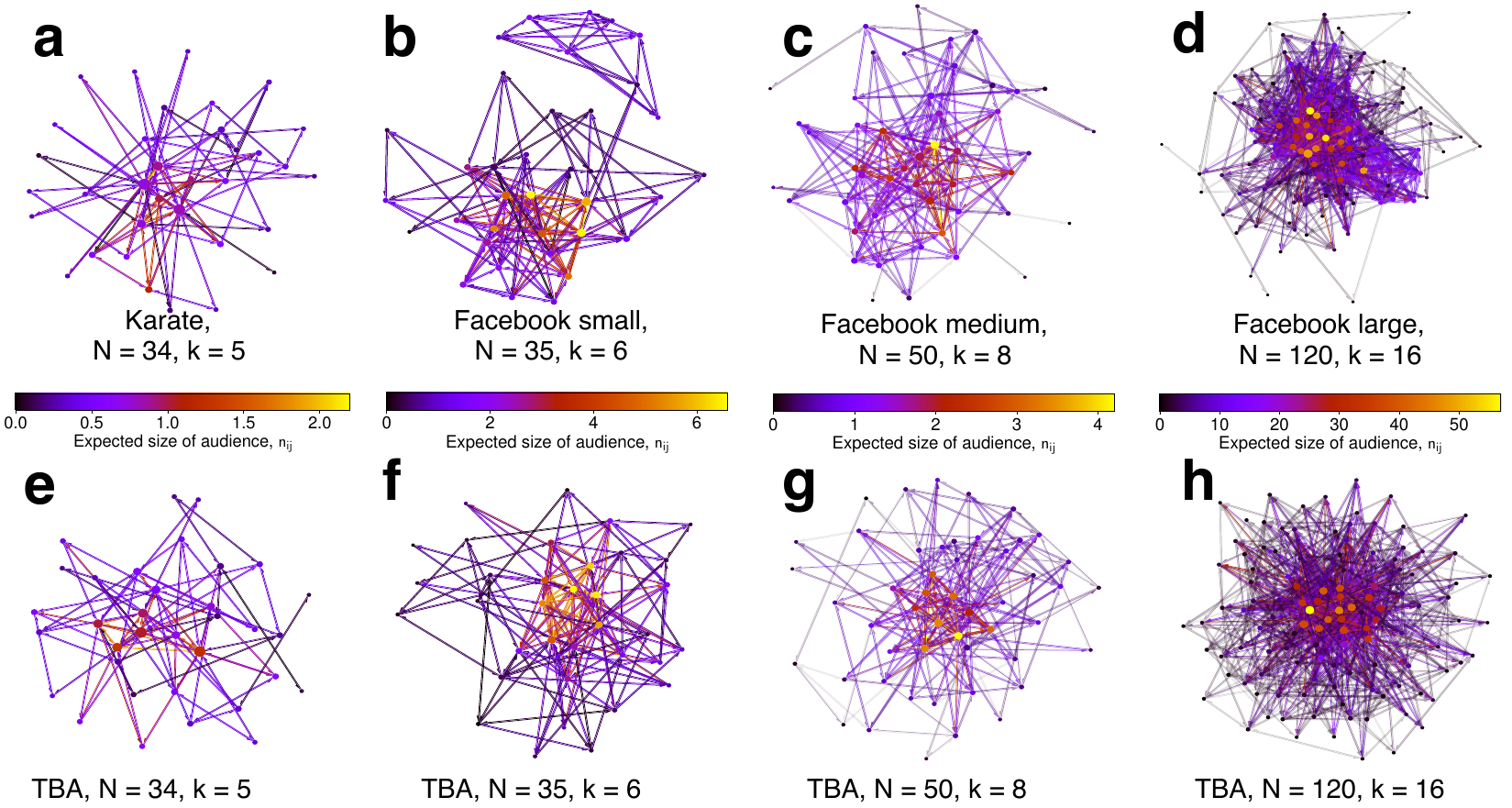}
}

\caption{{\bf Further examples of real and TBA generated networks}. The upper row shows various real-world networks (subsamples of the ``Facebook'' network, \cite{snapnets,leskovec2012learning} see ``Plotting social networks'' \SI{} for details), while the lower row shows networks generated by TBA with the same number of individuals $N$ and average degree $k$. Edges and nodes are colored as in {\bf Figure 4}. The real-world social networks and the TBA networks both show centrally and highly connected hubs, high clustering, as well as strong incentives throughout. The trust based attachment process generates highly cooperative populations that resemble real-world social networks in key ways.}
\label{fig:edfig6}
\end{figure*}

\begin{figure*}[p]
\noindent\makebox[\linewidth]{
  \includegraphics[width=0.8\linewidth]{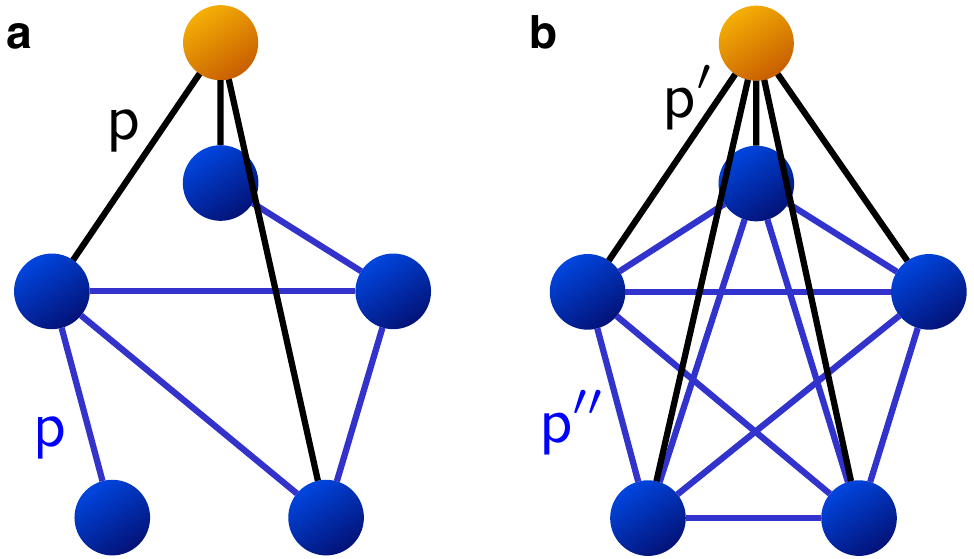}
}

\caption{{\bf Random local neighborhood approximation}. We show the actual local neighborhood {\bf(a)} of an actor node $i$ with $k = 6$ neighbors, where the recipient is shown in orange, and the other neighbors in blue. The actor itself is not shown. Every edge can transmit information with probability $p$. This is equivalent to keeping each edge with probability $p$ and transmitting information across the resulting network. The random graph approximation {\bf (b)} that these edges are randomly distributed. While this is true in expectation, it neglects the real structure of the local neighborhood. Mathematically, the assumption is that all edges from the recipient to the other neighbors (black edges) and all edges among the other neighbors (blue edges) exist at first. They are then kept with modified probabilities such that in expectation, the same number of edges from each category remains for information transfer as in the original graph. The sampling probability $p$ is modified for the black edges ($p'$) and the blue edges ($p''$) separately. In this case, we would have $p' = \frac{3}{5} p$ and $p'' = \frac{5}{10}p$, since on the real graph {\bf(a)}, only $3$ out of $5$ possible black edges exist, and $5$ out of $10$ possible blue edges. See the section ``Random neighborhood approximation for gossip propagation'' in the \SI{} for details.}
\label{fig:random_graph_approx}
\end{figure*}

\begin{figure*}[p]
\noindent\makebox[\linewidth]{
  \includegraphics[width=1.0\linewidth]{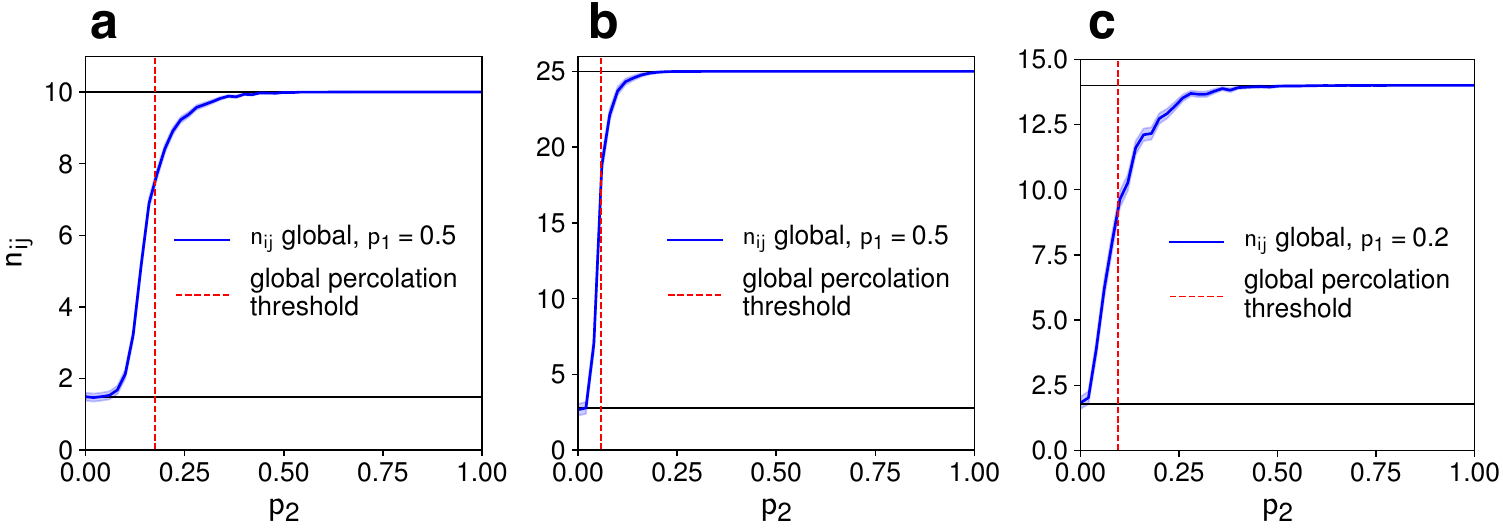}
}

\caption{{\bf Global percolation model}. We explore numerically the consequences of a model that allows for information flow on every edge of the actor's neighborhood network with probability $p_1$, and with probability $p_2$ on the rest of the network (excluding the edges originating from the actor themselves). In all panels, the value of $n_{ij}$ for a given actor and recipient resulting from this model is shown in blue, where the blue shaded region shows the numerical error (s.d.). The black horizontal lines give the value of $n_{ij}$ for the local model (lower line) and the maximum attainable value of $k-1$ (upper line). The vertical dashed line is the numerically determined edge percolation threshold on the overall network. {\bf(a)} A random actor and recipient are chosen from an Erdos-Renyi network of $N = 400$ nodes with an average degree of $10$. {\bf(b)} Same as {\bf(a)} but $N = 1000$ and an average degree of $30$. The larger, denser network leads to a sharper percolation transition. {\bf(c)} A random actor and recipient from a social network with $N = 4039$ individuals and an average degree of $44$. In all cases, the global percolation threshold $p_c$ determines the key regimes. For $p_2 < p_c$, the value of $n_{ij}$ approaches the value in the local model. For $p_2 > p_c$, $n_{ij}$ approaches $k-1$. For $p_2 \approx p_c$, there is a transition regime whose sharpness and exact behavior is determined by the overall network.}
\label{fig:global_percolation}
\end{figure*}



\newpage
\clearpage
\bibliography{citation}